\documentclass[12pt,a4paper]{article}
\pdfoutput=1
\usepackage{jheppub}
\usepackage{epstopdf}
\usepackage{graphicx}
\usepackage{epsfig}
\usepackage{dcolumn}  
\usepackage{bm}    
\usepackage{amssymb} 
\usepackage{amsmath,bm}
\usepackage{amsfonts}    
\usepackage{slashed}  
\usepackage[mathscr]{euscript}
\usepackage{epsfig}
\usepackage[position=t,singlelinecheck=off]{subfig}
\usepackage{caption}
\begin{document}
\def\be{\begin{equation}}
\def\ee{\end{equation}}
\def\bea{\begin{eqnarray}} 
\def\eea{\end{eqnarray}}

\title{Quasinormal modes and holographic correlators in a crunching AdS geometry}
\author{S. Prem Kumar}  
\author{and  Vladislav Vaganov}
\affiliation{
   Department of Physics, Swansea University, 
   Singleton Park, Swansea SA2 8PP, UK.
   }
\emailAdd{s.p.kumar@swansea.ac.uk, pyvv@swansea.ac.uk}
\date{\today}
\abstract{
We calculate frequency space holographic correlators in an asymptotically AdS crunching background, dual to a relevant deformation of the M2-brane CFT placed in de Sitter spacetime.  
For massless bulk scalars, exploiting the connection to a solvable supersymmetric quantum mechanical problem, we obtain the exact frequency space correlator for the dual operator in the deformed CFT. Controlling the shape of the crunching surface in the Penrose diagram by smoothly dialling the deformation from zero to infinity, we observe that in the large deformation limit the Penrose diagram becomes a `square', and the exact holographic correlators display striking similarities to their counterparts in the BTZ black hole and its higher dimensional generalisations. We numerically determine quasinormal poles for relevant and irrelevant operators, and find an intricate pattern of these in the complex frequency plane. In the case of relevant operators, the deformation parameter has  an infinite sequence of critical values, each one characterised by a pair of poles colliding and moving away from the imaginary frequency axis with increasing deformation. In the limit of infinite deformation all scalar operators have identical quasinormal spectra.
 We compare and contrast our strongly coupled de Sitter QFT results with strongly coupled thermal correlators from AdS black holes.}

\maketitle

\section{Introduction}
Quantum field theories (QFTs) in de Sitter (dS) spacetime are known and expected to exhibit certain features characteristic of field theories at finite temperature \cite{Birrell:1982ix, Spradlin:2001pw}. Whilst such physical aspects of de Sitter space QFTs have long been the subject of exploration in free or weakly interacting limits, the strong coupling limit throws up entirely new and  fascinating questions. Correlators of certain classes of large-$N$ thermal field theories at strong coupling in $d$-dimensions are computed holographically by black hole geometries in Anti-de Sitter (AdS) spacetimes in $d+1$ dimensions \cite{maldacena, witten}. Such Green's functions then exhibit frequency space non-analyticities in the form of simple poles at (complex) frequencies associated to quasinormal modes of the dual black hole geometries \cite{sonstarinets, starinets}. A remarkable feature of such strongly coupled thermal correlators is that they encode subtle signatures of the dual black hole singularity \cite{shenkeretal, fl1, fl2} in the complex frequency plane.  

It is natural to ask whether QFTs in de Sitter spacetimes share the above features of thermal correlators at strong coupling. Crucially, AdS/CFT duality relates strongly interacting non-conformal theories at large-$N$ and on (fixed) de Sitter spacetime to asymptotically AdS backgrounds with spacelike big crunch singularities \cite{maldacena1, harlow}. These crunches occur behind a horizon in the FRW patch of dS-sliced, asymptotically AdS geometries. A conformal transformation in the boundary QFT relates such setups to the so-called ``designer gravity'' models \cite{HH1, HH2} where the boundary QFT on the Einstein static universe is dual to an asymptotically global AdS geometry which evolves to a big crunch singularity in finite global time \cite{Barbon:2011ta, Barbon:2013nta}. 

Unlike the thermal CFT/black hole duality,  the correspondence between crunching AdS backgrounds and de Sitter space QFTs requires the field theories to be non-conformal i.e. obtained via relevant deformations of a conformal fixed point. The undeformed CFT on de Sitter space is dual to a non-singular geometry (AdS written in dS-sliced coordinates). Interesting physical phenomena can still ensue when one of the CFT coordinates is compactified and the bulk geometry exhibits phase transitions \cite{Hutasoit:2009xy, Marolf:2010tg, Balasubramanian:2002am}. Holographic correlators in these situations are relatively simple and exhibit some interesting features including thermal properties which closely resemble thermal effects in weakly interacting de Sitter QFTs.  However, the really interesting situations arise when CFT deformations are switched on and the bulk geometry has a curvature singularity. Various investigations of classes of such models have been undertaken with the goal of identifying holographic signatures of the bulk singularity and its potential resolution \cite{ Turok:2007ry, Craps:2007ch, EHH1, EHH2}\footnote{Related ideas have been explored in \cite{Das:2006dz,Craps:2006xq, Awad:2008jf}.}. 

Generally speaking, analytically tractable models of deformed CFTs are difficult to come by, and this is particularly important if the questions of interest involve analytic properties of correlation functions. In recent work \cite{paper1} we pointed out that a particular, single scalar truncation of ${\cal N}=8$ supergravity in four dimensions, with a known Euclidean solution \cite{Papadimitriou:2004ap, yiannis1, yiannis0, yiannis2}, provides (upon appropriate analytic continuation) an analytically tractable example of a gravity dual of a strongly coupled 
 QFT in de Sitter spacetime. The theory in question is the large-$N$ limit of the CFT on M2-branes \cite{maldacena, magoo} which is intrinsically strongly coupled, placed in three dimensional de Sitter space and deformed by a relevant operator with conformal dimension $\Delta=1$ at the conformal fixed point. The resulting gravity dual has several attractive features worthy of detailed exploration, some of which were already pointed out in \cite{paper1}. One of the properties of this system, which also happens to be a general feature of these types of crunching AdS models, is that geodesic probes which compute boundary correlators (of high dimension operators) stay away from, and therefore do not probe the bulk singularity \footnote{A very interesting way around this has  recently been argued in \cite{Bzowski:2015clm}.} \cite{hubeny, maldacena2, Fischler:2013fba, engelhardt}.
 
 The fact that geodesic limits of boundary correlators  stay away from the bulk singularity provides sufficient motivation to go beyond the limit of large dimension QFT operators. In particular, it may be argued that subtle signatures of black hole singularities are encoded in the locations of quasinormal poles in the complex frequency plane  \cite{fl1, hoyos}, 
  since the wave equation in the bulk, analytically continued into the FRW patch (where the crunch resides), should be sensitive to the geometry behind the horizon. With this motivation, we initiate a detailed study of frequency space correlators in the analytically tractable deformation of the M2-brane CFT,  described above. It is already quite interesting that the notion of a frequency space correlator can be made precise \cite{Hutasoit:2009xy} in  the de Sitter space QFT where time translation invariance is not manifest. In fact, for $s$-waves (or $\ell =0$ harmonics on the spatial sphere on the boundary), the temporal modes in $dS_3$ are simple exponentials and a frequency space Fourier transform is natural. We explore the frequency space analytic structure for relevant, irrelevant and marginal operators as a function of the CFT deformation parameter, and find several intricate features in each case, some of which provide tantalising hints of a signal of the bulk singularity. Our main results include the exact computation of the retarded Green's function for the $\Delta=3$ operator for any value of the CFT deformation, and  numerical determination of quasinormal poles in frequency space Green's functions for relevant and irrelevant boundary operators with arbitrary values of the deformation parameter. For the relevant operators (in particular $\Delta=5/2$) we uncover an extremely intricate behaviour of the quasinormal poles: poles collide and move off the imaginary axis in pairs as the CFT deformation is increased smoothly, resulting in an infinite (discrete) sequence of critical values for the deformation parameter, each associated to a particular pair of poles. For large enough deformations, the line of poles in the complex plane makes an angle whose value could be linked with the location of the crunch singularity in tortoise coordinates.  Curiously, in the limit of infinite deformation we find that the quasinormal spectra of all scalar operators appear to coincide.
  
  The paper is organised as follows: in sections 2 and 3, we summarise certain general aspects of crunching backgrounds, and basic features of the wave equation and associated potentials relevant for the holographic calculations of correlation functions. Section 4 provides a concise review of the deformed AdS$_4$ background which we study in this paper. Section 5, which forms the bulk of the paper contains all our results on quasinormal poles and correlation functions, and finally in section 6, we compare and contrast various aspects of the crunching AdS setup to the situation with AdS black holes. In appendix \ref{appa} we provide a brief derivation of the frequency space signals of the black hole singularity from the holographic viewpoint.

\section{Crunching backgrounds and AdS deformations}
Crunching AdS cosmologies dual to strongly coupled field theories in de Sitter spacetime can be obtained as Lorentzian continuations of  gravity duals of non-conformal field theories on spheres. A smooth asymptotically (Euclidean) AdS$_{d+1}$ geometry can be characterised by a metric of the form,
\be
ds^2\,=\,d\xi^2\,+\,a(\xi)^2\, d\Omega_{d}^2\,,\qquad 0 \leq \xi <\infty\,,
\ee
where $a(\xi)\sim e^{\xi}$ near the conformal boundary $(\xi \to\infty)$ which is the $d$-dimensional sphere $S^d$. Smoothness at the origin requires $a(\xi)$ to vanish as $a(\xi)\simeq \xi$  near $\xi=0$. Upon Wick rotation of the polar angle of the $S^d$ slice, the metric describes an asymptotically AdS geometry with de Sitter (dS$_d$) slices:
\be
ds^2\,=\,d\xi^2\,+\,a(\xi)^2\,\left(-dt^2\,+\,\cosh^2t\,d\Omega_{d-1}^2\right)\,.
\ee
The origin $\xi=0$ is now a horizon. The spacetime behind the horizon can be accessed by analytically continuing the coordinates $\xi$ and $t$:
\be
\xi\,\to\, i\sigma\,,\qquad  t\,\to\,\chi -\frac{i\pi}{2}\,,
\ee
with $\sigma,\chi \in {\mathbb R}$.  In this region the metric describes an FRW geometry with hyperbolic  (${\mathbb H}^d$) spatial slices, and an FRW scale factor given by 
\be
\tilde a (\sigma)\,\equiv\,-\,ia(i\sigma)\,.
\ee
The scale factor increases from zero at the horizon at $\sigma =0$, attains a maximum and vanishes a second time at $\sigma =\sigma_c$ where curvature invariants diverge and the point can be identified as a big crunch singularity. In the vicinity of the crunch, the scale factor vanishes as
\be
\tilde a(\sigma)\,\sim\, (\sigma-\sigma_c)^\gamma\,, \label{powerlaw}
\ee
with $\gamma < 1$ generically. In two related works \cite{paper1, paper2} in the case of three specific microscopic models with $d=3$ and 4 we find that the exponent $\gamma = 1/d$. In this article we will focus our attention on the AdS$_4$  deformation (discussed in \cite{paper1}) which happens to be analytically tractable. Certain qualitative features of this model, however, appear to be generic. 

\section{Scalar wave equation}

The correlator of a scalar operator ${\cal O}_\Delta$ in the boundary de-Sitter-space field theory is obtained by examining the solutions to the wave equation for a dual bulk field of mass $m_\varphi$, with the standard relationship between $m_\varphi$ and the conformal dimension $\Delta$ of the (UV) CFT operator,
$\Delta \,=\,\frac{3}{2}+\sqrt{m^2_\varphi +\frac{9}{4}}$ (assuming $m_\varphi^2 >-2$).
The isometries of the de Sitter slices  allow a natural separation of variables. Following the line of reasoning adopted in \cite{Hutasoit:2009xy, paper1}, we may expand the (probe) bulk field $\varphi$ in terms of temporal de Sitter harmonics ${\cal T}_\ell (\omega, t)$, and spatial spherical harmonics on $S^{d-1}$:
\be
{\varphi}(\xi, t,\Omega)\,\sim\,\sum_{\ell, \,m}A_{\ell m} Y_{\ell m}(\Omega)\int\frac{d\omega}{2\pi}\,{\varphi}_{\omega}(\xi)\,{\cal T}_\ell(\omega, t)\,.
\ee
The harmonics ${\cal T}_\ell$ are solutions to a temporal differential equation with a  P\"oschl-Teller potential and can be chosen so that  the late time and/or high frequency limits yield plane wave type solutions ${\cal T}_\ell (\omega, t) \to e^{-i\omega t}$. This depends on whether $d$ is even or odd, as discussed in more detail in the appendix of \cite{paper1}. 
 
 Below, we choose to focus attention on harmonics with $\ell =0$ or $s$-waves, and for this reason we suppress the $\ell$-index entirely. The equations of motion for the harmonics $\varphi_\omega$ with $\ell=0$ can then be used to  compute $s$-wave correlators of the boundary operator ${\cal O}_\Delta$. After a field rescaling,  
\be
\psi_\omega\equiv a^{(d-1)/2}\,\varphi_\omega\,,
\ee
and transforming to tortoise coordinates, 
\be
z\,=\,\int_{\xi}^\infty\frac{d\zeta}{a(\zeta)}\,,
\ee
the harmonics $\psi_\omega(z)$, satisfy a Schr\"odinger-like wave equation 
\be
-\frac{d^2\psi_\omega}{dz^2}\,+\,V(z)\,\psi_\omega(z)\,=\,\omega^2\,\psi_\omega(z)\,.
\ee
The Schr\"odinger potential $V(z)$ can be expressed in terms of the scale factor $a(z)$ and its derivatives,
\bea
V(z)&=&V_0(z)\,+\, V_1(z)\,,\qquad V_0\,=\,\,m_\varphi^2\,a^2
\,,\\\nonumber
V_1(z)&=&(d-1)\frac{a''}{2a}\,+\,{(d-1)(d-3)}\left(\frac{a'}{2a}\right)^2\,-\,\tfrac{1}{4}(d-1)^2\,.
\eea
The potential has certain generic features and some specific ones which have been found to hold across different examples \cite{paper2}. On general grounds $V(z)$ decays exponentially whilst approaching the bulk horizon, $z\to \infty$ and diverges as $V(z)\sim z^{-2}$ near $z=0$, the conformal boundary.

Continuing to the FRW region behind the horizon, which is achieved by $z\to w-\frac{i\pi}{2}$ with $w \in {\mathbb R}$, the potential diverges at the crunch $(z=z_c)$ as $V(z)\sim (z-z_c)^{-2}$. When the exponent in \eqref{powerlaw} satisfies $\gamma < 2/(d+1)$, the potential term is driven to negative infinity. This is the case for examples studied in \cite{paper1, paper2}. 

\begin{figure}
\centering
\includegraphics[width= 2.4in]{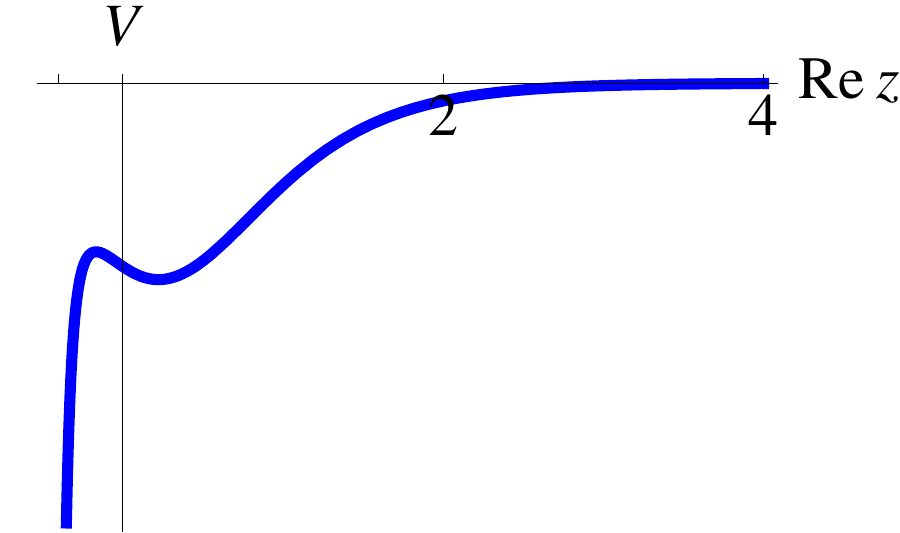}\hspace{1.0in}
\includegraphics[width= 2.4in]{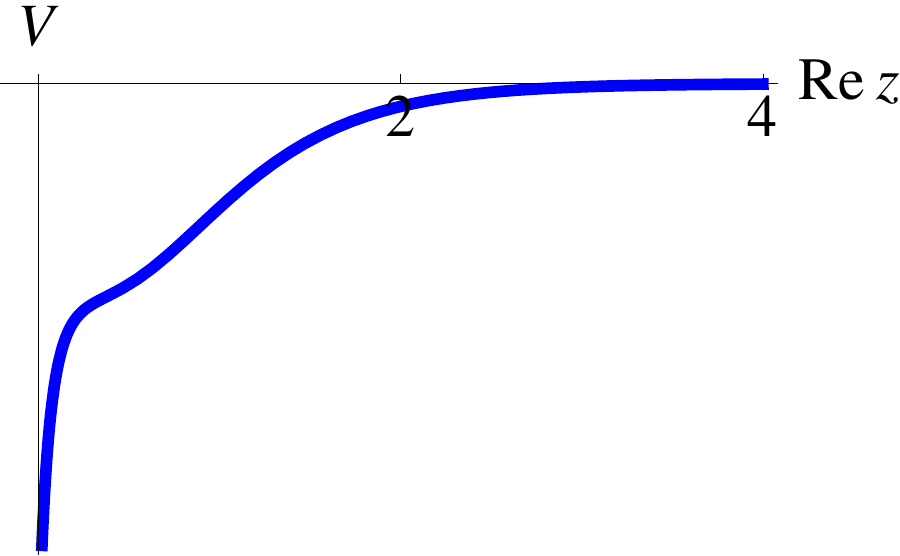}
\caption{\small{The effective Schr\"odinger potential in the probe scalar wave equation in the FRW region (${\rm Im}(z)=-\frac{i\pi}{2}$) for sufficiently small (left) and large (right) CFT deformation.}}\label{shape}
\end{figure}

For the AdS$_4$ deformation that we focus on in this paper \cite{paper1} , $\gamma = 1/3$ and the Schr\"odinger potential diverges as $V(z)\sim -\frac{1}{4}(z-z_c)^{-2}$ at the crunch singularity. The potential has two additional features stemming from a competition between $V_0$ and $V_1$. Since the scale factor increases and then decreases to zero in the FRW patch, the contribution $V_0$ is non-monotonic behind the horizon and vanishing at the crunch whilst $V_1$ diverges at the singularity. This competition implies that the shape of the potential in the FRW patch depends on the strength of the CFT deformation, as illustrated in 
figure \ref{shape}.
As the magnitude of the CFT deformation is increased, the Schr\"odinger potential continued into the FRW patch undergoes a qualitative transition from being non-monotonic with two extrema prior to the crunch, to a monotonically decreasing function diverging at the crunch. 

\section{An AdS$_4$ deformation}
The exact solution for the deformation of Euclidean  AdS$_4$ which we make use of can be found in the works \cite{Papadimitriou:2004ap, yiannis0, yiannis1, yiannis2}. It arises from a particular single scalar truncation of $\mathcal{N}=8$ gauged supergravity in four dimensions, distinct from the truncation  considered in \cite{HH1, HH2}.   The boundary quantum field theory is most naturally viewed as a relevant deformation of the M2-brane CFT formulated on $S^3$.
Setting the AdS radius to unity, the system has a one-parameter\footnote{The parameter $f_0$ in this paper differs slightly from that used in \cite{paper1}: in particular, $(f_0)_{\rm here}\,=\,(f_0/\sqrt{6})_{\rm there}$.} family of regular solutions with $S^3$ slices:   
\bea
&&ds^2 \,=\, \left(1-f(u)^2\right)\left[\frac{du^2}{u^2(1+u^2)}\,+\,\frac{1}{u^2}\, d\Omega_3^2\right]\,,\label{defads}  \\\nonumber \\\nonumber
&&\Phi \,=\, \sqrt{6}\tanh^{-1}{f(u)}\,,\\\nonumber
&&f(u)\,=\, \frac{f_0 \,u}{\sqrt{1+u^2}+u\sqrt{1+f_0^2}}\,.
\eea
The scalar $\Phi$ has mass squared $M^2=-2$ in units of the AdS radius. As this lies within the window $-9/4 < M^2 < -5/4$, there exist two interpretations of the associated deformation, determined by choice of boundary conditions \cite{kw}.  The constant $f_0$ controls the strength of the deformation and can naturally be interpreted as a deformation by a relevant operator of dimension $\Delta=1$, which is one of the viewpoints suggested for instance in \cite{maldacena1}. The asymptotic expansion of $\Phi(u)$ near the boundary $u=0$ yields,
\be
\frac{1}{\sqrt 6}\Phi(u)\,=\,f_0 u \,-\,f_0\sqrt{1+f_0^2}\,u^2 \,+\ldots
\ee
which allows us to read off the strength of the $\Delta=1$ deformation ($f_0$) and its VEV from the coefficients of $u^2$ and $u$, respectively.

Upon analytic continuation to Lorentzian signature, the background describes a strongly interacting quantum field theory placed in dS$_3$ spacetime whose gravity dual has a bulk horizon at $u=\infty$. 
As is customary, it is useful for this exercise to pass to the tortoise coordinate which is non-singular at the horizon ($u=\infty$) in the Lorentzian continuation of \eqref{defads}:
\bea
&& z\,=\,\sinh^{-1}u\,,\qquad f_0\,=\,{\rm cosech}\, z_0\,,\label{zaz}\\\nonumber\\\nonumber
&& ds^2\,=\,a^2(z)\,\left(dz^2\,-dt^2\,+\,\cosh^2t\,d\Omega_2^2\right)\,,
\\\nonumber\\\nonumber
&&a^2(z)\,=\,\sinh^{-2}z\,-\,\sinh^{-2}(z+z_0)\,,
\eea
where the coordinate range is $0 < z <\infty$ with $z=0$, the conformal boundary of AdS$_4$. In this parametrisation, the situation with vanishing deformation $(f_0\to 0)$ corresponds to $z_0 \to \infty$ (keeping $z$ fixed), while an infinitely large deformation is obtained as $z_0$ approaches the AdS boundary $z_0 \to 0$.

\begin{figure}
\centering
\includegraphics[width= 1.9in]{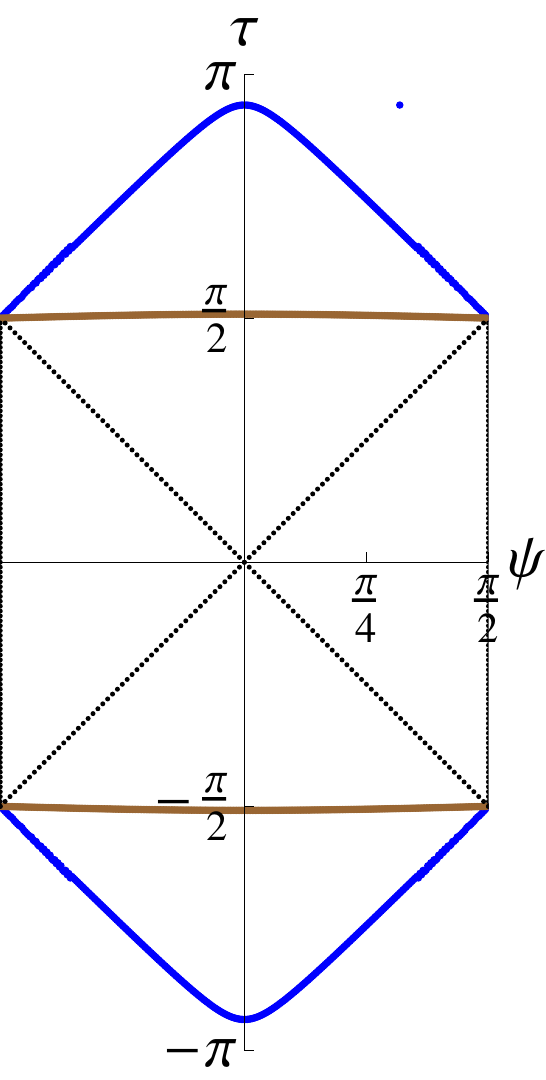}
\caption{\small{The shape of the FRW patch singularities for $f_0=0.02$ (blue) and $f_0=20$ (brown) in the deformed AdS$_4$ geometry. In the limit of large deformations the Penrose diagram is a square. The radial coordinate $\psi$ ranges between $0$ and $\frac{\pi}{2}$. Unlike the AdS black hole, there is only one conformal boundary, and the diagram is really a solid cylinder with caps. The left and right sides of the figure above should be viewed as antipodal points on the spatial $S^2$ section of the boundary. }}\label{crunchshape}
\end{figure}

The spacetime behind the bulk horizon can be accessed by analytically continuing the coordinates $z$ and $t$:
\be
z\to -\frac{i\pi}{2}\,+\,w\,,\qquad t \to \chi\,-\,\frac{i\pi}{2}\,,
\ee
with $w,\chi \in {\mathbb R}$, so that
\be
ds^2\,=\,\tilde a (w)^2\,(-dw^2 \,+ \,d\chi^2\,+\,\sinh^2\chi \,d\Omega_2^2 )\,,
\ee
where $i\tilde a (w)\,=\,a(z)$.
In the FRW patch, the scale factor is vanishing at the horizon ($w\to \infty$) and at the big crunch which occurs when
\be
z\,=\,z_c\,\equiv -\frac{z_0}{2}\,-\,\frac{i\pi}{2}\,.
\ee
The qualitative difference between the Penrose diagrams, specifically with regard to the shape of the crunch, for small and large deformations, is shown in figure \ref{crunchshape}. The global coordinates $(\tau,\psi)$ are defined via the relations
\be
\tan\frac{\tau-\psi}{2}\,=\,-e^{-t-z}\,,\qquad \tan\frac{\tau+\psi}{2}\,=\,
e^{t-z}\,.
\ee
For small $f_0$, the crunching surface lies close to the null cone $\tau +\psi\,=\,\pi$ deviating from it near $\psi=0$, while for large $f_0$, the crunching surface approaches $\tau = \frac{\pi}{2}$ and the Penrose diagram becomes a square.

\section{Quasinormal modes and holographic correlators}
The connection between the analytic structure of the frequency space correlator we compute below and its real time behaviour is clearest for $\ell=0$ modes i.e. excitations that are homogeneous along the spatial section of the boundary de Sitter geometry. This is because the  temporal modes ${\cal T}_\ell (t)$ which are given by the associated Legendre polynomials,
\be
{\cal T}_\ell(t)\,=\,\frac{\Gamma(1-i \omega)}{\cosh t}\,P_\ell^{i\omega}(\tanh t)\,,
\ee
setting aside the kinematic redshift factor, become pure exponentials when $\ell=0$:
\be
{\cal T}_0(t)\,=\,\frac{e^{i\omega t}}{\cosh t}\,.
\ee
The redshift factor cancels against a corresponding contribution that appears in the measure of the action for the free scalar $\varphi$ in the dS-sliced bulk. Then, following the standard steps for calculating 
position-space holographic correlation functions, the retarded frequency space correlator can be defined as the ordinary (temporal) Fourier transform of the real time retarded Green's function:
\bea
&& G_R(t,t')\,=\,-i\theta(t-t')\int \frac{d\Omega}{4\pi}\int\frac{d\Omega'}{4\pi}\,\langle[{\cal O}_\Delta(\Omega,t),{\cal O}_{\Delta}(\Omega',t')]\rangle\,,\\\nonumber\\\nonumber
&& \tilde G_R(\omega)\,=\,\int_{-\infty}^\infty dt\,e^{-i\omega(t-t')}\,G_R(t,t')\,.
\eea
Importantly, the $s$-wave real time correlator turns out to be a function of $(t-t')$ only because the temporal modes are effectively pure exponentials (see also \cite{Hutasoit:2009xy}).

Singularities of frequency space Green's functions determine the non-trivial real time behaviour of correlation functions. At strong coupling, thermal field theory correlators computed holographically using AdS black hole backgrounds exhibit poles in the complex frequency plane which correspond to quasinormal modes of the black hole geometry. For AdS black holes, it can be argued that high frequency quasinormal poles reflect the location of the singularity behind the horizon \cite{fl1, hoyos}.  In particular, the values of the higher quasinormal frequencies are determined by the complex  Schwarzschild time  taken by a null geodesic to go from the AdS boundary to the singularity. It was argued in \cite{hoyos} that this is due to  null cone singularities in boundary correlators, which could be identified in a geometrical optics or eikonal approximation as arising from null rays reflecting off the black hole singularity and the AdS boundary, thus bouncing  around the Penrose diagram. 
 
With this motivation, we  proceed to examine the quasinormal frequencies and the frequency space (retarded) correlator following from the scalar wave equation in the deformed AdS$_4$ geometry \eqref{defads}. The wave equation for the radial mode of a bulk scalar with mass $m_\varphi$ and frequency $\omega$ in Schr\"odinger form is,
\bea
&&-\psi_\omega''(z)\,+\,V(z)\,\psi_\omega(z)\,=\,\omega^2\,\psi_\omega(z)\,,
\label{waveqn}\\\nonumber\\\nonumber
&&V(z)\,=\,\frac{m^2_\varphi+2}{\sinh^2 z}\,-\,
\frac{m^2_\varphi-2}{\sinh^2 (z+z_0)}\,-\,\frac{1}{\sinh^2(2z + z_0)}\,.
\eea
On the cylinder $- \frac{\pi}{2}\leq{\rm Im} (z) < \frac{\pi}{2}$, the  potential has (regular) singular points at $z=0$, the AdS boundary, and the crunch singularity in the FRW patch at 
\be
z_c\,=\, -\frac{z_0}{2}-\frac{i\pi}{2}\,.\label{zc}
\ee
There are two additional double poles in the unphysical domain $z<0$, at $z=-z_0$ and $z=-z_0/2$.


\begin{figure}
\centering
\includegraphics[width= 3.8in]{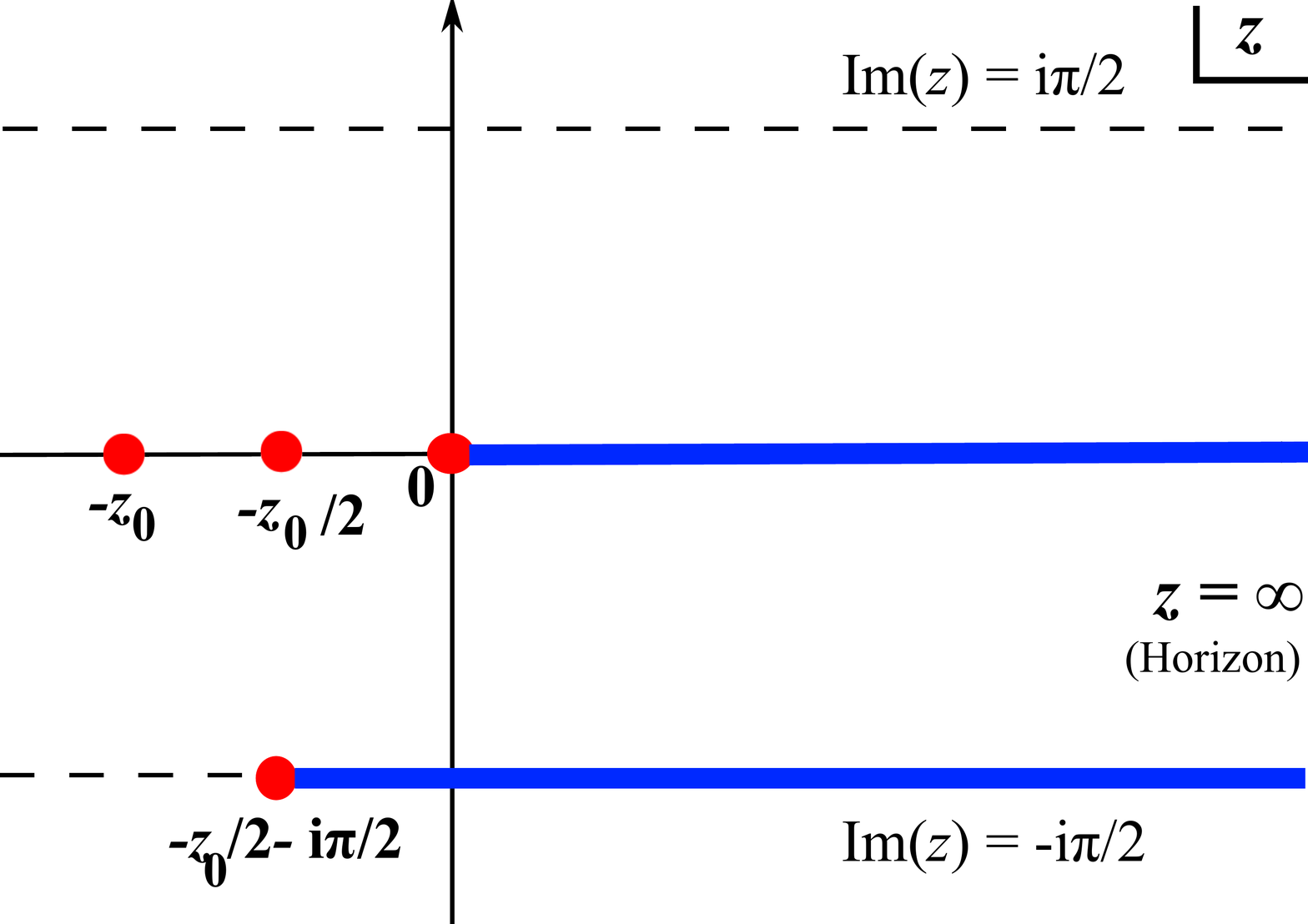}
\caption{\small{The poles of $V(z)$ in the domain $-\frac{\pi}{2} \leq {\rm Im}(z) < \frac{\pi}{2}$ in the $z$-plane. The physical regions are indicated in blue: the asymptotically AdS exterior $0< z<\infty$ and the interior FRW patch along ${\rm Im}(z)=-\frac{\pi}{2}$ in the range ${\rm Re}(z_c) < {\rm Re }(z) < \infty$, where $z_c=-\frac{i\pi}{2} - \frac{z_0}{2}$ is the location of the crunch singularity. The pole at $z=0$ corresponds to the AdS boundary while two other poles appear in unphysical regions at $z=-z_0/2$ and $z=-z_0$.}}\label{zplane}
\end{figure}

The quasinormal poles are found by demanding a purely infalling wave at the horizon, 
\be
{\psi}_\omega(z)\,\sim\, e^{i\omega z}\,,\qquad z\to \infty\,,
\ee
accompanied by a normalisable boundary condition at the AdS boundary at $z=0$:
\be
\psi_\omega(z)\,\sim \, z^{\frac{1}{2}+q}\,,\qquad q\equiv \sqrt{\frac{9}{4}+m_\varphi^2}\,.
\ee
Alternatively, in analytically tractable situations, one may calculate the frequency space retarded correlator using the Son-Starinets prescription \cite{sonstarinets} and look for its singularities (in the lower half complex $\omega$-plane). Our analysis will comprise of the following exact analytical and numerical results:
\begin{itemize}
\item{Exact analytical expression for the retarded  Green's function for vanishing deformation and any mass $m_\varphi$.}
\item{Exact result for the retarded correlator with $m_\varphi=0$ and arbitrary deformation}
\item{Numerical determination of the quasinormal frequencies for arbitrary mass and deformation.}
\end{itemize}

\subsection{Zero deformation}
The undeformed geometry being AdS$_4$ (in the dS-sliced description), all correlators can be calculated analytically. The limit $f_0\to 0$ corresponds to taking $z_0\to \infty$. The retarded Green's function in frequency space is readily calculable as in \cite{paper1} and is given by
\be
{\tilde G}_{R}(\omega)\,=\,4q\,2^{-2q}\,\frac{\Gamma(-q)\,\Gamma\left(\tfrac{1}{2}+q-i\omega\right)}{\Gamma(q)\,\Gamma\left(\tfrac{1}{2}-q-i\omega\right)}\,,\label{undeformed}
\ee
where $q=\sqrt{m_\varphi^2+9/4}$. It is analytic in the upper half plane and has simple poles in the lower half plane at,
\be
\omega\,=\,-i\left(\frac{1}{2}+q+n\right)\,,\qquad n=0,1,2,\ldots
\ee
The quasinormal  poles for general $m_\varphi$ are integer spaced in pure (dS-sliced) AdS$_4$ spacetime. 

The spectrum of chiral (protected) scalar operators in the M2-brane CFT with conformal dimensions $\Delta$, dual to bulk scalars with masses $m_\varphi$, is given by \cite{magoo}
\be
m_\varphi^2\,=\,\frac{1}{4}k(k-6)\,,\qquad \Delta\,=\,\frac{k}{2}\,,\qquad
k=2,3,\ldots.
\ee
Thus $q$ is a half-integer (integer) when $\Delta$ is integral (half-integral). A straightforward evaluation of \eqref{undeformed} at integral or half-integral values of $q$ can be misleading as the resulting expressions are polynomials in $\omega$. Instead, one must obtain these via a limiting process, making note of the fact that on ${\mathbb R}^3$, the frequency space Green's function $\tilde G_R(\omega)\sim \omega^{2\Delta-3}\ln\omega^2$ for half-integral values of $\Delta$ and 
$\tilde G_R(\omega)\sim \omega^{2\Delta-3}$ when the conformal dimension $\Delta$ takes any other value (see e.g. Appendix A of \cite{sonstarinets} for an analogous situation in four dimensions where only operators with {\em integer} dimensions have $\tilde G_R(\omega)\sim \omega^{2\Delta-4}\ln\omega^2$).

Therefore, for example, in the case of an operator with $\Delta=5/2$, we expand \eqref{undeformed} in a Taylor series about $q=1$ and keep the leading non-analytic piece:
\be
\tilde G_R(\omega)\left.\right|_{\Delta\to 5/2}\,=\,-\frac{1}{2}(4\omega^2+1)\,\Psi\left(-\tfrac{1}{2}-i\omega\right)\,+\, {\rm analytic}\,.\label{5/2}
\ee
Here $\Psi(z)\,=\,\Gamma'(z)/\Gamma(z)$ is the digamma function and we have omitted polynomials in $\omega$ for the sake of clarity. The digamma function $\Psi(z)$ has simple poles with unit residue at $z=0$ and the negative integers $z=-n$ (with $n\in {\mathbb Z}$). Therefore, the quasinormal poles of \eqref{5/2} appear at $\omega_n=-\left(\frac{3}{2}+n\right)i$ with $n=0,1,2,\ldots$. In the high frequency limit, using $\Psi(z)\simeq \ln z$ for $z\gg 1$, we immediately obtain $\tilde G_R(\omega)\sim \omega^2 \ln\omega$, as expected on general grounds. On the other hand, operators with integer $\Delta$ have regular frequency space Green's functions.

\subsection{Massless scalar for any deformation}
When the CFT deformation is non-vanishing it is not {\it a priori} obvious that the wave equation can actually be solved analytically. For the special case of a massless bulk scalar $m_\varphi=0$ dual to a marginal operator ($\Delta=3$) in the boundary UV theory, the equation obeyed by the radial mode $\psi_\omega(z)$ can be solved exactly for any value of the deformation of parameter. This is possible due to the fact that the massless radial  equation is related in a simple way to a supersymmetric quantum mechanical system. To see this, we first define 
\be
W(z)\,\equiv\,\frac{a'(z)}{a(z)}\,,\qquad {\cal D}_z\,\equiv \partial_z \,+\, W(z)\,.
\ee
Then the massless radial equation can be expressed as
\be
{\cal D}_z{\cal D}_z^\dagger\,{\psi}_\omega^{+} \,=\,(\omega^2+1)\,\psi_\omega^{+}.
\ee 
The Schr\"odinger potential for this equation is $V^{+}(z)\,=\, W'(z)\,+\,W(z)^2\,=\,a''(z)/a(z)$.
The solutions to this equation can be inferred from the so-called isospectral problem, namely 
\be
{\cal D}^\dagger_z{\cal D}_z\,{\psi}_\omega^{-} \,=\,(\omega^2+1)\,\psi_\omega^{-}\,,
\ee
which has a completely different Schr\"odinger potential $V^{-}(z)\,=\,W(z)^2\,- \,W'(z)$. The associated Hamiltonians are often referred to as bosonic and fermionic partner Hamiltonians (see e.g. \cite{Cooper:1994eh}) and are isospectral and further possess closely related reflection and transmission coefficients. It follows immediately that the wave functions $\psi_\omega^{(\pm)}$ are related as
\be
\psi_\omega^{+} \,\sim\,{\cal D}_z\,\psi^{-}_\omega\,,\qquad
\psi_\omega^{-} \,\sim\,{\cal D}_z^\dagger\,\psi^{+}_\omega\,. \label{relations}
\ee
For our deformed AdS$_4$ geometry, the original wave equation \eqref{waveqn} has the Schr\"odinger potential $V^{+}(z)$ which is not obviously integrable. However, the partner potential, 
\be
V^{-}(z)\,=\,1+ \frac{3}{\sinh^2(2z+z_0)}\,,\label{partnerpot}
\ee
which yields the Schr\"odinger-like equation,
\be
\left[-\partial_z^2\,+\,3\,{\rm cosech}^{2}(2z+z_0)\right]\,\psi^{-}_\omega(z)\,=\,\omega^2\,\psi_\omega^{-}(z)\,,\label{partnerpot}
\ee
is the so-called P\"oschl-Teller potential and is exactly solvable in terms of associated Legendre polynomials. Using this solution and the relations \eqref{relations} 
 we determine the exact retarded Green's function after implementing the Son-Starinets prescription. In terms of the solution $\psi_\omega^-$ to the P\"oschl-Teller problem, the holographic retarded Green's function can be expressed compactly in terms of the wave function for the partner potential \eqref{partnerpot}
\be
{\tilde G}_{R}(\omega)\,=\,-2(\omega^2 + 1)\,\frac{\psi_\omega^{-'}(0)}{\psi_\omega^-(0)}\,,
\ee
where we have suitably adjusted the overall multiplicative normalisation constant to match the results of the undeformed AdS background, as explained below.
Picking $\psi_\omega^-$ such that it yields an infalling wave at the horizon, we have
\be
\psi_\omega^-(z)\,=\,P_{1/2}^{i\omega/2}\left[\coth (2z+z_0)\right]\,.
\ee
Therefore the frequency space retarded correlator for a massless probe field in the deformed AdS$_4$ geometry is
\be
\boxed{\left.{\tilde G}_{R}(\omega)\right|_{\Delta=3}\,=\,-2(\omega^2 + 1)\,\partial_z\ln\left.\left(P_{1/2}^{i\omega/2}\left[\coth (2z+z_0)\right]\right)\,\right|_{z=0}}\,.\label{exactret}
\ee 
\subsubsection{Small deformation limit}
We now examine the exact result in the limit of small deformation ($f_0\ll 1 $ or equivalently $z_0\gg 1$). In the undeformed AdS$_4$ background the strict massless limit $m_\varphi \to 0$ of the correlator in eq.\eqref{undeformed} yields
\be
{\tilde G}_{R}\,\to\, 2i(\omega^3 + \omega)\,,
\ee  
which is a curious result since it is regular and therefore only yields contact terms as a function of time.  This limit is reproduced by the exact formula \eqref{exactret}, which reveals a somewhat more intricate picture. In particular, expanding for large $z_0$ (or $f_0\ll 1$) we find
\be
{\tilde G}_{R}(\omega)\,=\,2i(\omega^3 + \omega)\,+\,\frac{3i\omega + \coth z_0}{\sinh^2z_0}\,+\,
\sum_{n=1}^\infty\frac{c_n}{(\omega + 2n\,i)+\delta_n}
\ee
where (using $f_0={\rm cosech} z_0$ ) the coefficients $c_n$ and $\delta_n$ can each be expanded in a powers series in $f_0$ 
\bea
&&c_1\,=\,9i\,f_0^2 + \ldots\,,\qquad
c_2\,=\, \tfrac{225}{16}\,i\,f_0^4 + \ldots\,,\qquad c_n\,\sim\, {\#}i f_0^{2n}+\ldots
\\\nonumber\\\nonumber
&&\delta_1\,=\,\tfrac{3}{8}i f_0^2+ \ldots\,,\qquad
\delta_2\,=\,\tfrac{15}{256}\,i\,f_0^4+\ldots\,,\qquad \delta_n\,\sim\,{\#}i f_0^{2n}+\ldots\,
\eea 
\begin{figure}
\centering
\includegraphics[width= 3.0in]{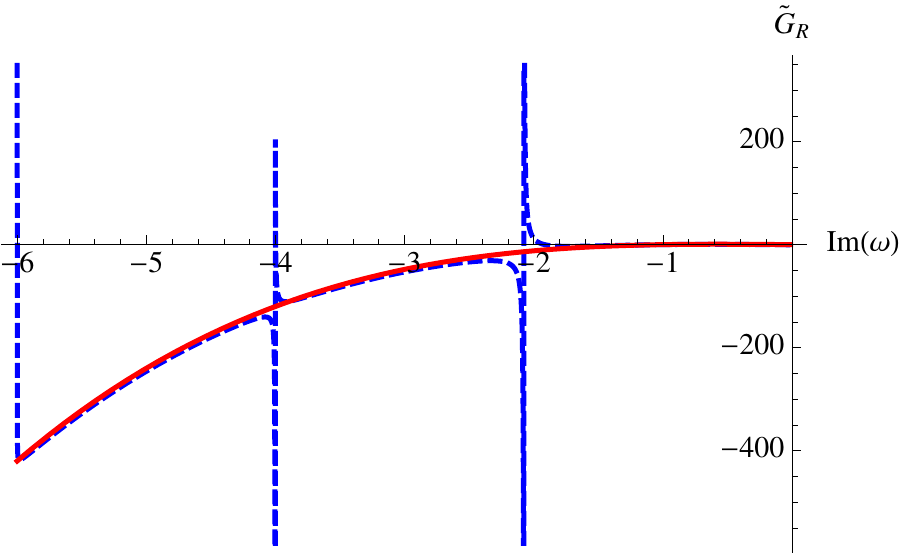}
\caption{\small{The retarded Green's function $\tilde G_R(\omega)$ for the $\Delta=3$ operator plotted (blue, dashed curve) along the negative imaginary frequency axis for $f_0=0.45$ which captures the small deformation limit surprisingly well. The quasinormal poles are close to $\omega_n \simeq - 2 n i$ for $n=1,2,\ldots$. Plotted in red is the undeformed case ($f_0=0$).
 }}\label{weakpole}
\end{figure}
Therefore, an infinitesimal deformation produces an infinite number of quasinormal poles along the negative imaginary axis located at $\omega_n\simeq - 2n i$. This is also confirmed by the numerical plot of the correlation function shown in figure \ref{weakpole}. It is noteworthy that the residues of the poles at $\omega=\omega_n$ scale as $f_0^{2n}$ for small deformations, so that the effect of progressively higher quasinormal poles is parametrically suppressed.
 It would be very interesting to understand the precise physical origin of this effect.  The most interesting finding is that an infinitesimal deformation $f_0\ll 1$ which introduces a bulk crunch singularity, also simultaneously induces quasinormal pole singularities in the correlator for the massless (minimally coupled) bulk scalar.

\subsubsection{Infinite deformation}
For arbitrarily large deformations ($z_0=0$ or $f_0\to \infty$), the exact expression \eqref{exactret} for the frequency space correlator of the  operator with $\Delta=3$ (note this is the scaling dimension in the undeformed (UV) CFT), yields a simple limiting form in terms of the digamma function:
\bea
\left.{\tilde G}_{R}(\omega)\right|_{f_0 \to \infty}\,&&\to\,\label{infinitef0}\\\nonumber
&&(1+\omega^2)\left[6f_0+ \frac{2 i}{f_0}(-3\omega+2i)+\frac{3}{f_0}
(1+\omega^2)\left(\Psi\left(\tfrac{3}{2}-\tfrac{i\omega}{2}\right)\,-\,\ln(2f_0)\right)\right]\,.
\eea
Hence the quasinormal frequencies in the limit of infinite deformation are located at 
\be
\omega_n\,=\,-(2n+1)i\,,\qquad n=1,2,\ldots,
\ee
therefore determined by the {\em odd} integers as opposed to the 
small deformation limit when the poles on the negative imaginary axis are fixed by the even integers. \begin{figure}
\centering
\includegraphics[width= 2.7in]{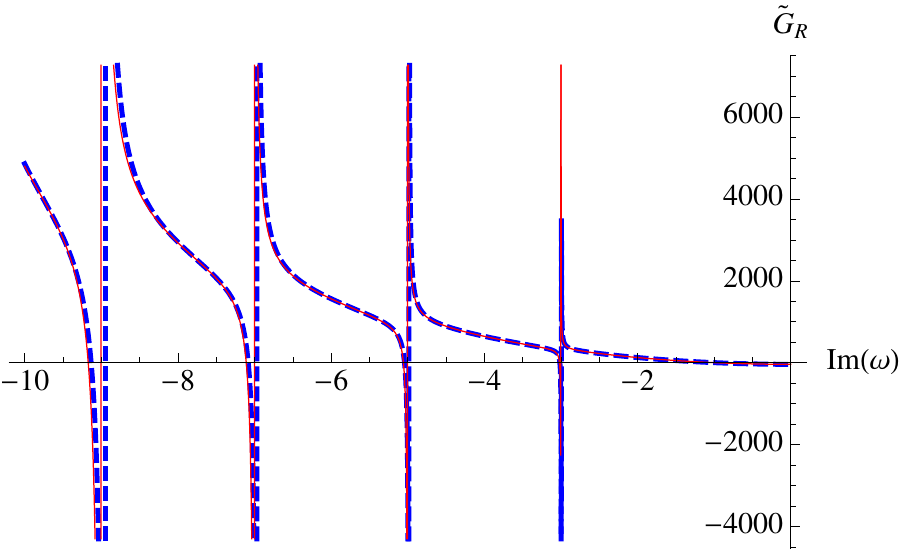}\hspace{0.4in}
\centering
\includegraphics[width= 2.8in]{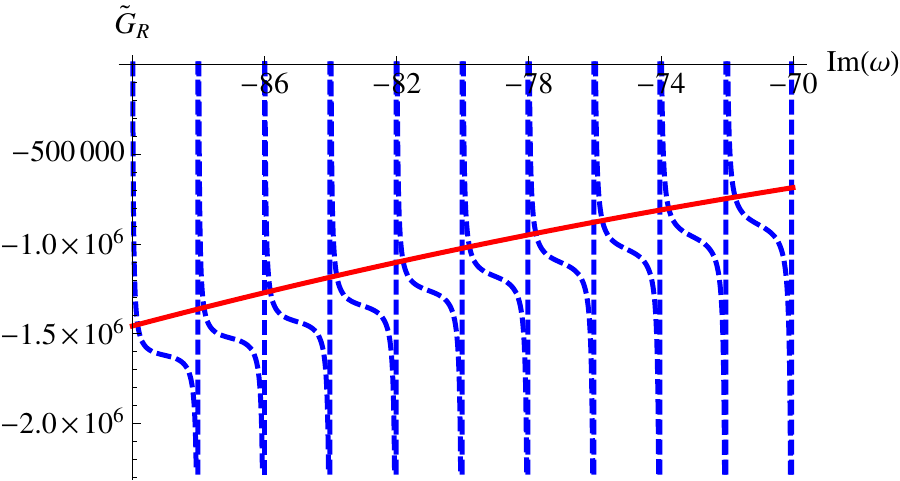}
\caption{\small{{\bf Left:} The retarded Green's function $\tilde G_R(\omega)$ for  $\Delta=3$ (blue, dashed curve) along the negative imaginary frequency axis for large deformation $f_0=20$. The quasinormal poles are close to $\omega_n \simeq - (2 n+1) i$ for $n=1,2,\ldots$. Plotted in red and practically indistinguishable from the blue curve is the limit of infinite deformation given by eq.\eqref{infinitef0}.
{\bf Right:} At high frequencies $\tilde G_R(\omega)$ with $f_0=20$ reverts to pure AdS-like behaviour with the poles determined by {\em even integers} superimposed on  an envelope $\sim 2i(\omega^3 +\omega)$ (solid red) which characterises  the undeformed AdS limit.}}\label{strongpole}
\end{figure}
In contrast to the limit of zero deformation where the strengths of the poles are parametrically suppressed, the  residues at the quasinormal poles  in the limit of large deformation are of the same order in $1/f_0$.

The correlator with $f_0\gg 1$ has another important feature. The leading non-analytic term in the $1/f_0$-expansion \eqref{infinitef0}, at large frequencies scales as $\sim \omega^4 \ln \omega^2$ which is the scaling expected from an operator with $\Delta=7/2$ in three dimensions. On the other hand, for {\em fixed} large $f_0$, the asymptotic high frequency limit reverts to that of ordinary AdS with $\tilde G_R \sim \omega^3$. This situation, clarified in figure \ref{strongpole},
 indicates a flow from the UV fixed point to an IR scaling regime when the deformation $f_0$ is arbitrarily large. The large-$f_0$ geometry 
 is determined by the scale factor $a(z)^2$:
 \be
 a(z)^2\,\simeq\,\frac{2}{f_0}\frac{\cosh z}{\sinh^3 z}\,,\qquad z \gg \frac{1}{f_0}\,,\qquad f_0 \gg 1\,.
 \ee
 This is not AdS spacetime, and so whilst there is no IR fixed point, high frequency correlation functions display an intermediate scaling behaviour dictated by this IR geometry.
 
In the limit of infinite deformation the Penrose diagram is a square (see figure \ref{crunchshape}), or more accurately, a solid cylinder with flat caps, reminiscent of the BTZ black hole and its higher dimensional generalisations, sometimes called ``topological AdS black holes'' \cite{Banados:1998dc}. The topological AdS and BTZ  black holes are locally AdS 
spacetimes with certain global identifications, and their respective singularities are associated to the shrinking of a spatial circle\footnote{The field theory dual to the topological black hole in $AdS_{d+1}$ spacetime is a CFT on $dS_{d-1}\times S^1$ for large enough $S^1$.}.

Our setup is a deformation of AdS$_4$ with a bulk curvature singularity and hence, qualitatively distinct from topological AdS black holes. It is therefore extremely interesting that when the Penrose diagram in the large deformation limit is essentially identical to BTZ and its higher dimensional generalisations, the detailed form of the correlator \eqref{infinitef0} and its singularities becomes strikingly similar to those computed in  BTZ \cite{sonstarinets} and the topological   AdS black holes \cite{Hutasoit:2009xy}. This similarity includes the appearance of the digamma function with equally spaced poles along the negative imaginary axis determined by  {\em odd} integers. 

\subsection{Quasinormal poles for $m_\varphi^2 \neq 0$}

When $m_\varphi^2\neq 0$, the scalar wave equation is not analytically solvable. Therefore we focus our attention on the quasinormal poles which can be determined numerically. We build the solution to the bulk wave equation in a Frobenius expansion around the horizon \cite{Horowitz:1999jd}, and then impose the requirement that the solution is normalizable near the boundary.  The numerical evaluation is performed by expressing $\psi_\omega$ as a truncated power series in the variable $y\equiv e^{-2z}$:
\be
\psi_{\omega,N}(y)\,=\,y^{i\omega/2 -1}\left(1+\sum_{n=1}^N\, c_n(\omega)\,y^n\right)\,,\qquad \qquad 0 < y <1\,.\label{frobenius}
\ee
where $N$ is sufficiently large ($\sim 10^2$), and the coefficients 
$c_n(\omega)$ are completely determined via recursion relations that follow from the Schr\"odinger-like equation for $\psi_\omega$.
Here $y=0$ is the horizon and $y=1$, the conformal boundary. 

Near the conformal boundary ($y\to 1$ or $z\to 0$) the mode $\psi_\omega$ has the two possible behaviours $\sim z^{\frac{1}{2}\pm q}$. We restrict our attention to the range $q>\frac{1}{2}$ which translates to $m_\varphi^2>-2$, so that only the asymptotics with $\psi_\omega \sim z^{\frac{1}{2}+q}$ is permitted for a normalisable solution. In particular, for $m_\varphi^2>-2$, requiring $\psi_\omega$ to vanish at the conformal boundary at $y=1$ (or $z=0$) automatically picks out the normalizable  solution.  Imposing this on the truncated expansion $\psi_{\omega,N}$ we are led to the condition 
\be
\psi_{\omega,N}(y=1)\,=\,1+\sum_{n=1}^N{c_n(\omega)}\,=\,0\,.
\ee 
The equation can now be seen as an order $N$ polynomial in $\omega$ whose roots yield the $N$ lowest quasinormal poles. The method gives accurate results for the lower quasinormal modes, improving with increasing $N$.

\begin{figure}
\centering
\includegraphics[width= 3.8in]{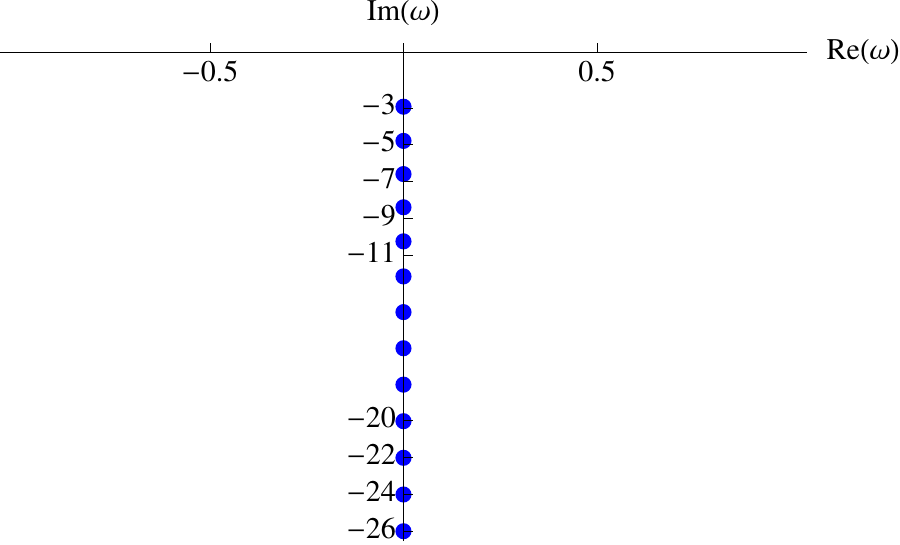}
\caption{\small{The quasinormal poles for $m_\varphi^2=0$ ($\Delta=3$) determined numerically for $f_0\,=\,5.45$. The lowest poles are determined by odd integers, whilst the high frequency poles match the even integers (AdS limit), as expected from the exact Green's function.}}
\label{qnmzero}
\end{figure}

\paragraph{{Check for $m_\varphi^2=0$:}} Implementing this method for the massless case, which we solved exactly in the previous section, serves as a quick check of the numerics. As $f_0$ is increased we expect to see quasinormal poles at $\omega_n=-(2n+1)i$ with $n\in {\mathbb Z}$. In particular this should apply for the lowest lying poles initially, whilst the high frequency poles should be dictated by the even integers as for pure AdS. This is precisely what figure \ref{qnmzero} shows for $f_0\simeq 5.45$.

\subsection{Quasinormal poles for $-2<m_\varphi^2<0$}
Next, we examine the quasinormal poles associated to operators that are relevant at the conformal fixed point, so that $m_\varphi^2<0$ for the dual bulk field. For simplicity, we further restrict attention to values $m_\varphi^2>-2$, which means that the bulk scalar has a unique normalizable mode. This places the conformal dimension in the range $2 < \Delta < 3$, and for the M2-brane CFT, picks out $\Delta=5/2$. This restriction is unnecessary if we view the bulk scalar simply as a probe and choose any value of $m_\varphi^2$ corresponding to the range $2<\Delta<3$. Given that there is no qualitative difference in the results, we therefore pick $m_\varphi^2=-5/4$ (equivalently, $\Delta=5/2$) as a representative case which also fits the spectrum of the M2-brane CFT.

\begin{figure}
\centering
\includegraphics[width= 1.9in]{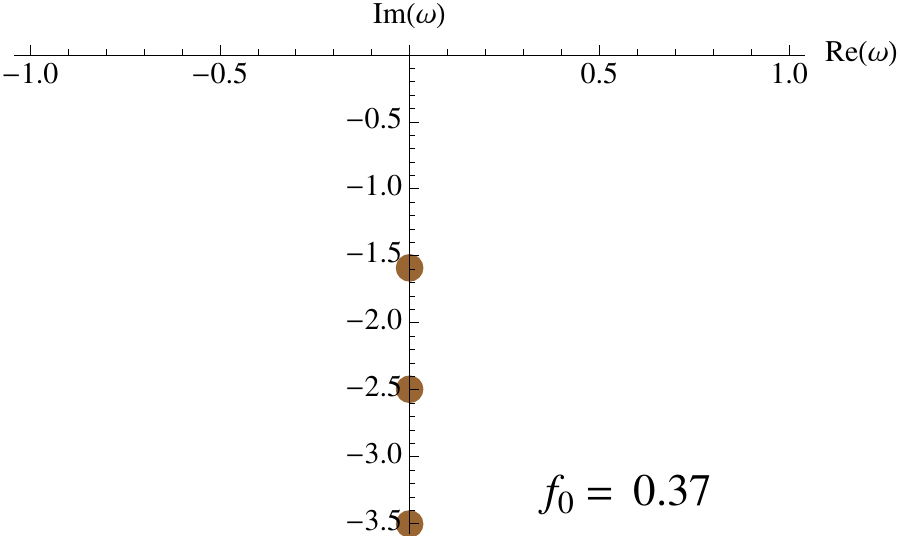}
\includegraphics[width= 1.9in]{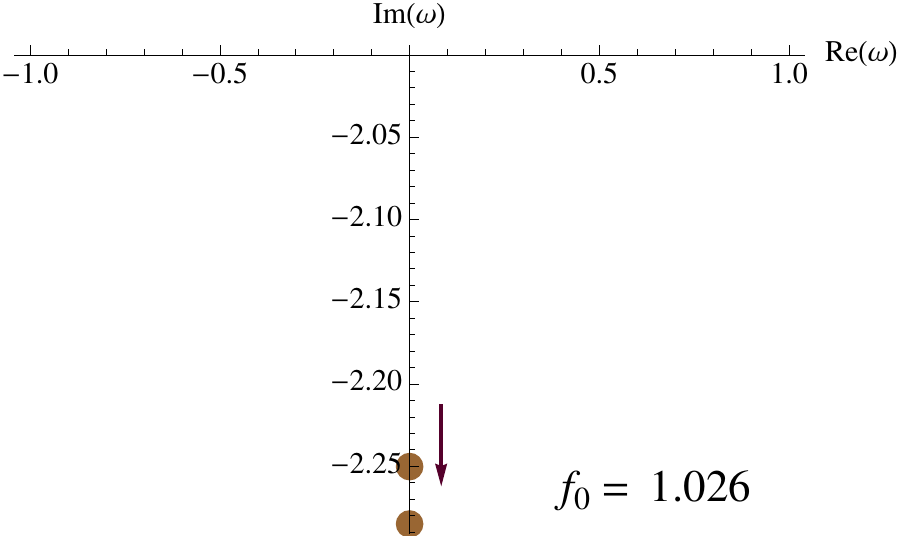}
\includegraphics[width= 1.9in]{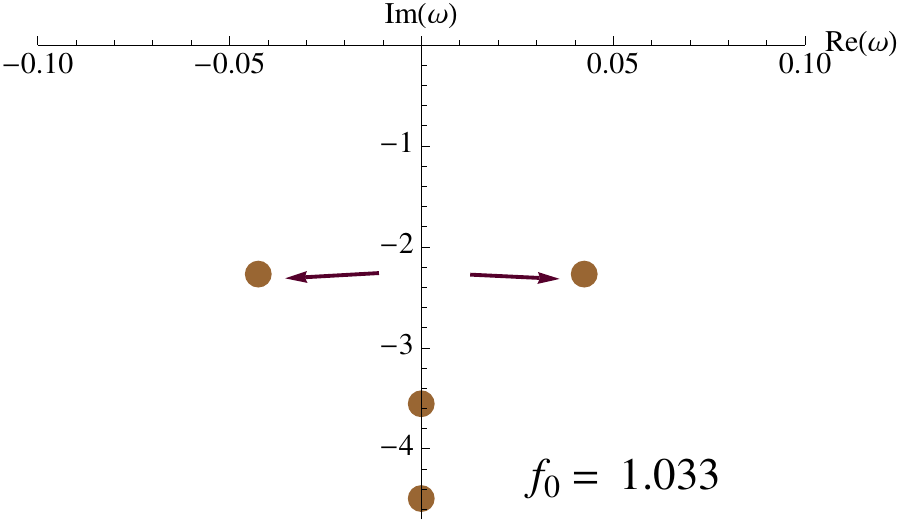}
\caption{\small{Merger and movement of lowest pair of quasinormal poles into complex plane beyond $f_0\approx 1.027$, for the $\Delta=5/2$ operator.}}
\label{qnmlow}
\end{figure}

Precisely at $f_0=0$, i.e. in the undeformed theory, we know from eq.\eqref{5/2} that the quasinormal poles are at 
\be
\omega_{n}\,=\,-\left(\tfrac{3}{2}+n\right)i\,,\qquad f_0=0\,,\qquad n=0,1,2 \ldots\,
\ee
As $f_0$ is smoothly dialled from zero we see the first signs of an interesting phenomenon which repeats with increasing $f_0$: the lowest quasinormal pole (originally at $\omega_1=-1.5 i$) moves toward and eventually merges with the second lowest pole ($\omega_2\simeq -2.5 i$ for small $f_0$) when the deformation parameter hits the critical value $f_0\approx 1.027$. Beyond this critical value of $f_0$, these two poles move into the complex plane. This sequence of events is shown in figure \ref{qnmlow}. Dialling $f_0$ further we find that the phenomenon repeats for every successive pair of poles. The next pair of poles $\omega_3$ and $\omega_4$, which start off at $-3.5 i$ and $-4.5i$, respectively, in the undeformed theory, merge and then become complex pairs beyond $f_0\approx 2.14$. 

\begin{figure}
\centering
\includegraphics[width= 5.8in]{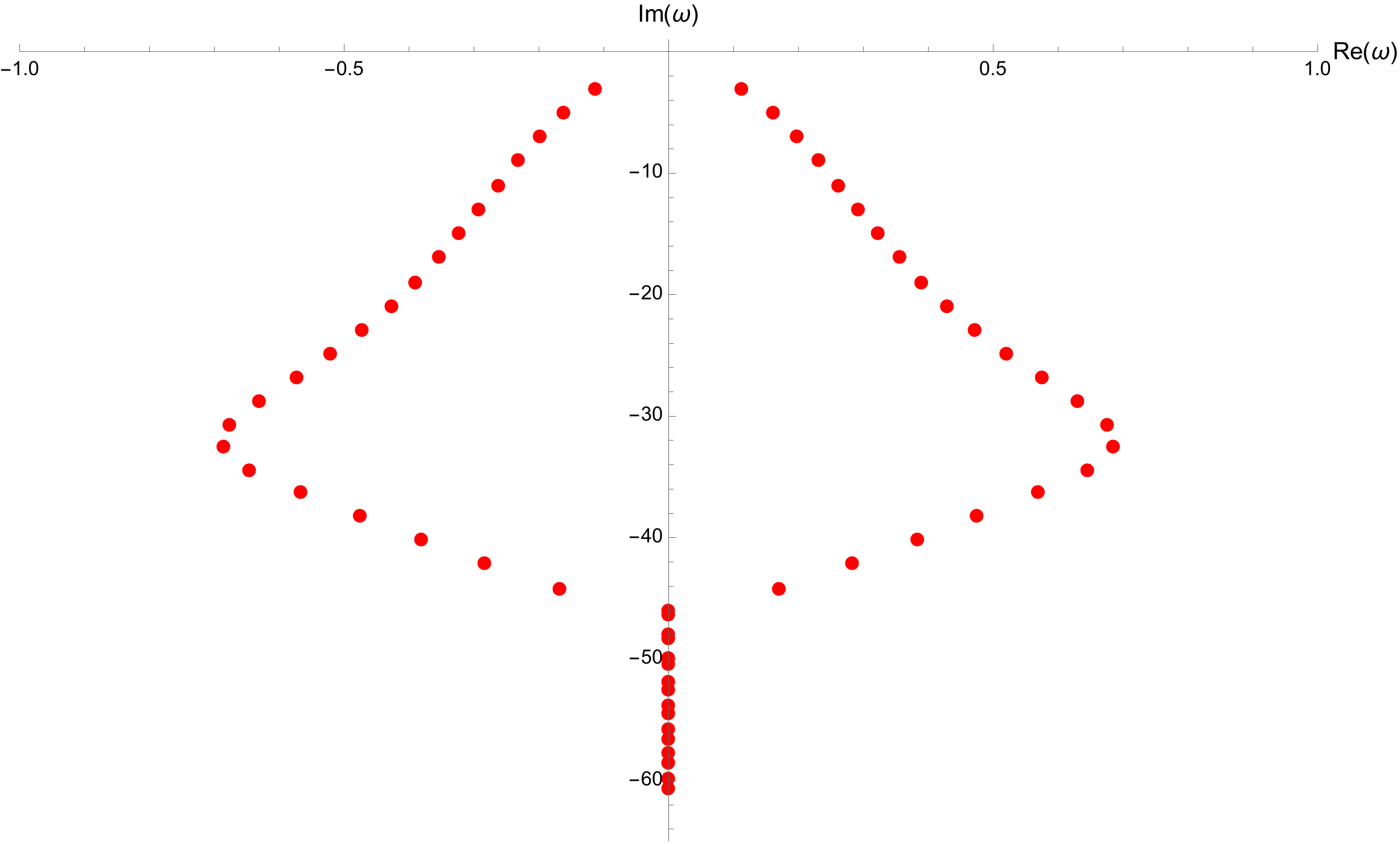}
\caption{\small{The quasinormal poles corresponding to the $\Delta=5/2$ operator ($m_\varphi^2= -5/4$) determined numerically for $f_0\,\simeq\,25$. Upon reaching the imaginary axis, the poles rapidly revert to pure AdS behaviour at high frequencies.}}
\label{qnm2}
\end{figure}

\begin{figure}
\centering
\includegraphics[width= 3.5in]{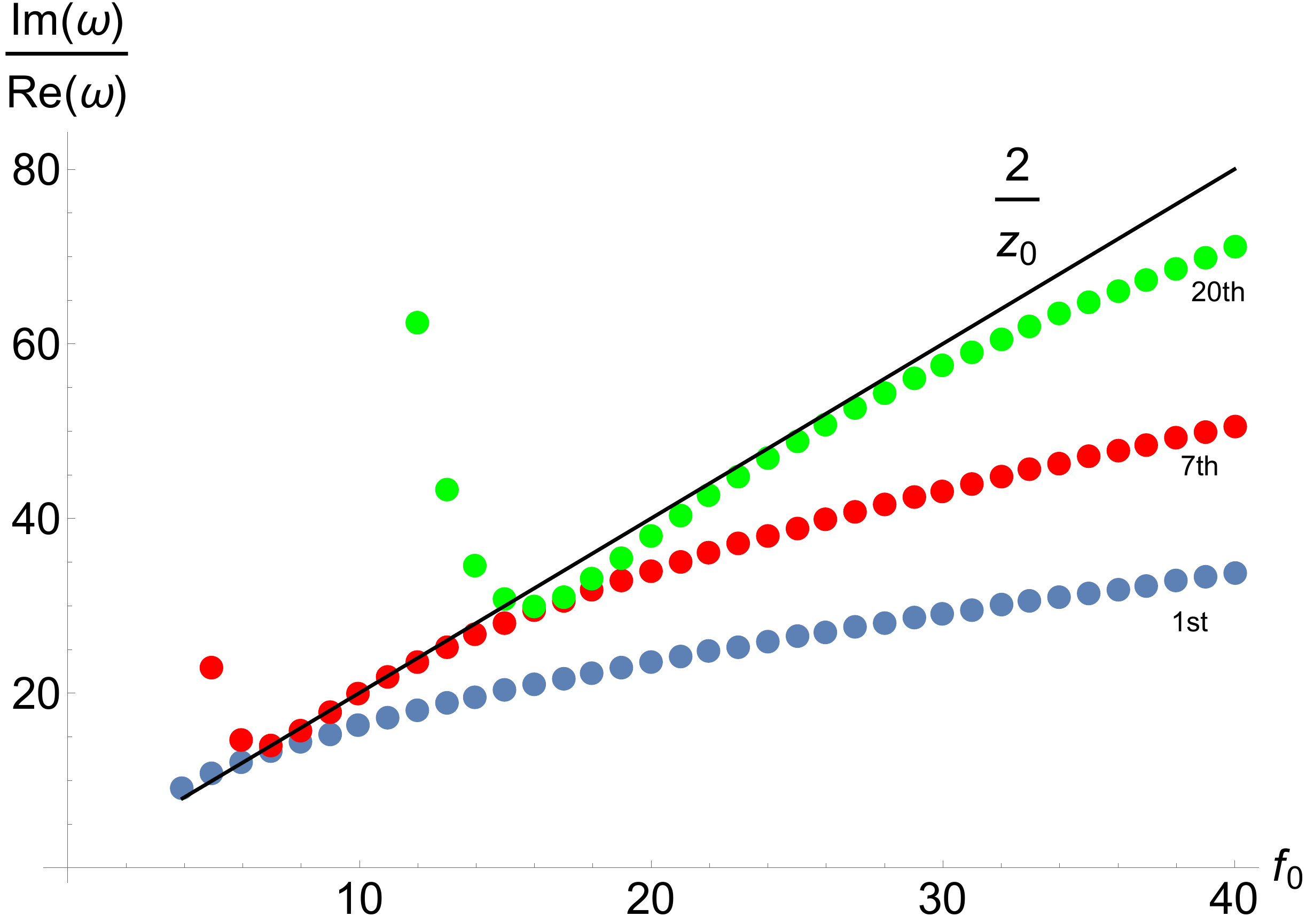}
\caption{\small{The angle made by the line of quasinormal poles (first, seventh and twentieth shown above) with the imaginary axis approaches a limiting value at large $f_0$, for  poles located away from the ends of the distribution in the complex plane.}}
\label{qnmangle}
\end{figure}

The motion of pairs of poles into the complex plane eventually yields an intricate pattern  for large $f_0$, as shown in figure \ref{qnm2}.
The poles that remain on the imaginary axis revert to AdS-like behaviour at high frequencies. At sufficiently large $f_0\gg 1$, we empirically confirm that the poles that move off the imaginary axis have
\be
{\rm Re}({\omega_n})\neq 0\,,\qquad {\rm Im}(\omega_n)\,\simeq\,-(2n+1)i\,,\qquad n=1,2,\ldots\,,\qquad f_0\gg 1\,.
\ee
\begin{figure}
\centering
\includegraphics[width= 4.5in]{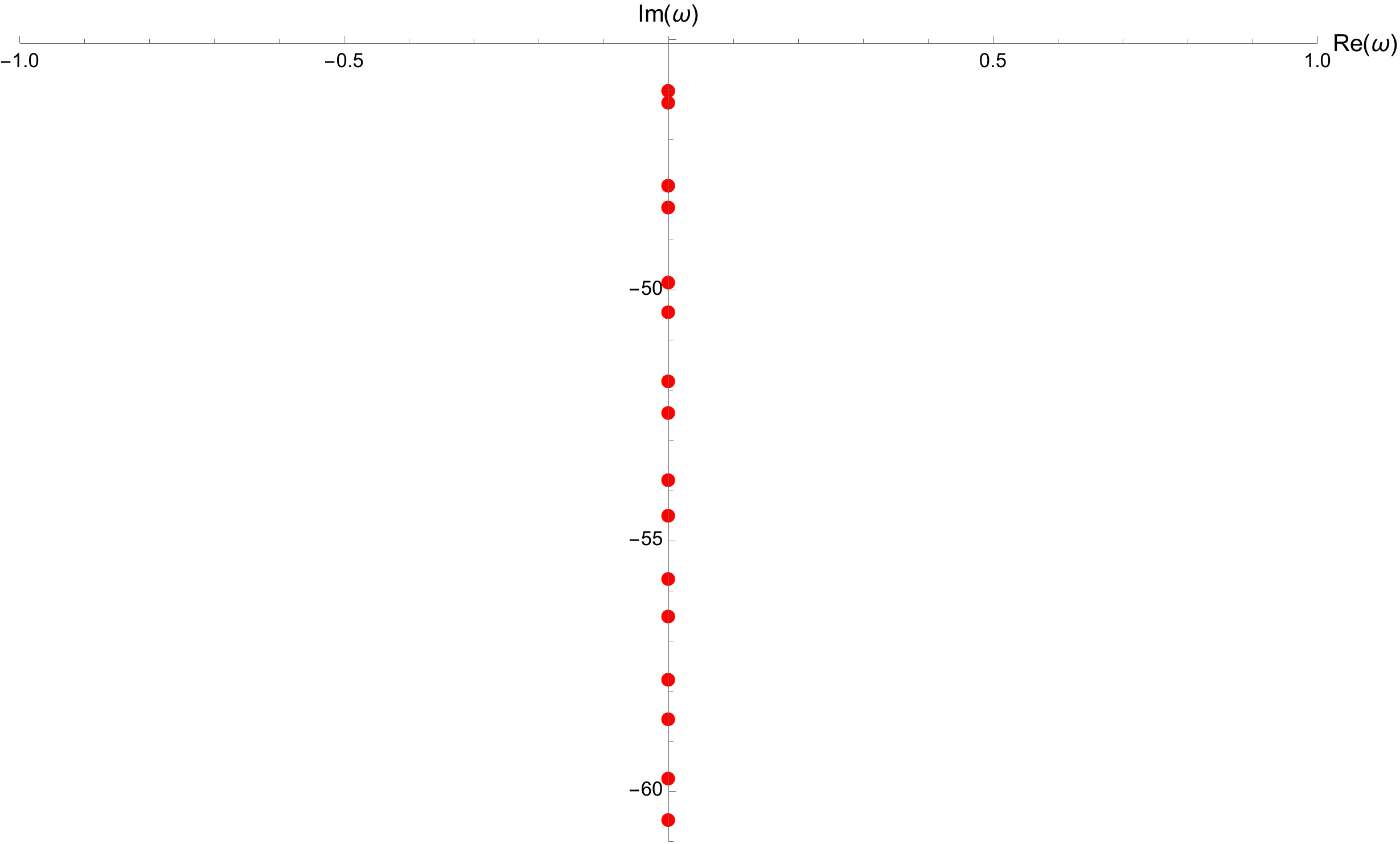}
\caption{\small{The high frequency quasinormal poles corresponding to the $\Delta=5/2$ operator at $f_0\,\simeq\,25$, gradually approaching the expected AdS-like behaviour with poles at the half-integers (see eq.\eqref{5/2}).}}
\label{qnm3}
\end{figure}
The real parts of the quasinormal frequencies  in the complex plane are not immediately obvious, particularly due to the effects near the low end of the quasinormal spectrum and at the higher end where the distribution of poles veers toward the imaginary axis. However as $f_0$ is increased, most of the poles remain away from the edges of the distribution and we can evaluate their real parts numerically. Shown in figure \ref{qnmangle} are the ratios of the real and imaginary parts of the quasinormal frequencies as a function of the deformation $f_0$. 
Based on the numerical results we infer that for sufficiently large deformations the locations of the higher quasinormal modes away from the imaginary axis is given to a good approximation by the simple formula
\be
\omega_n^\pm\approx\, (2n+1)\left(\pm\frac{z_0}{2}\,-\, i\right)\,,\qquad\qquad
f_0\gg1\,, \qquad n>>1\,,
\ee
where $n$ is in an integer and the superscripts $\pm$ refer to modes on either side of the imaginary axis.  A particularly tantalising aspect of this behaviour is that the location of the crunch singularity in tortoise coordinates is given by $z_c=-z_0/2-i\pi/2$, so that we may write
\be
\omega_n^\pm\,\approx\, (2n+1)(\pm {\rm Re}(z_c)-i)\,.
\ee
Furthermore, in the large $f_0$ limit $z_0\approx\frac{1}{f_0}$. Therefore the simple poles $\omega_n^\pm$ approach the imaginary axis in pairs and merge in the infinite deformation limit at $\omega_n= -(2n+1)i$. 

The dependence of the quasinormal poles on the crunch coordinate $z_c$ is intriguing and we will return to discuss the potential significance of this in the next section.

\subsection{Quasinormal poles for $m_\varphi^2>0$}
 
 In the case of bulk modes with $m_\varphi^2>0$, corresponding to irrelevant operators in the UV CFT, the numerical analysis of the quasinormal poles reveals a simple and interesting picture. The poles are always situated on the negative imaginary axis for all values of $f_0$. In the limit of undeformed AdS we know the exact result from eq.\eqref{undeformed}. In particular, in AdS spacetime, the lowest quasinormal  frequency is at $\omega = -i\left(\sqrt{m_\varphi^2 +\tfrac{9}{4}}+\tfrac{1}{2}\right)$, which, for large enough masses becomes $\omega \approx - im_\varphi$.

\begin{figure}
\centering
\includegraphics[height= 1.8in]{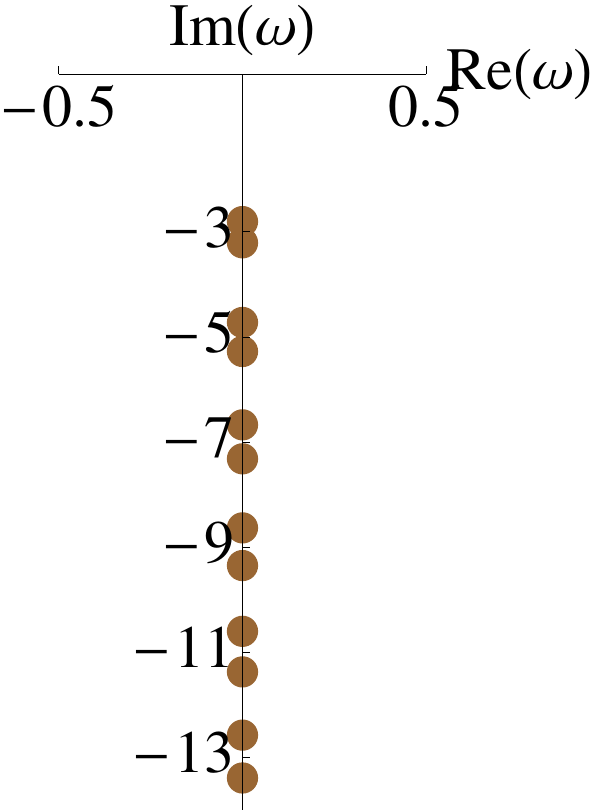}\hspace{0.8in}
\includegraphics[height = 1.8in]{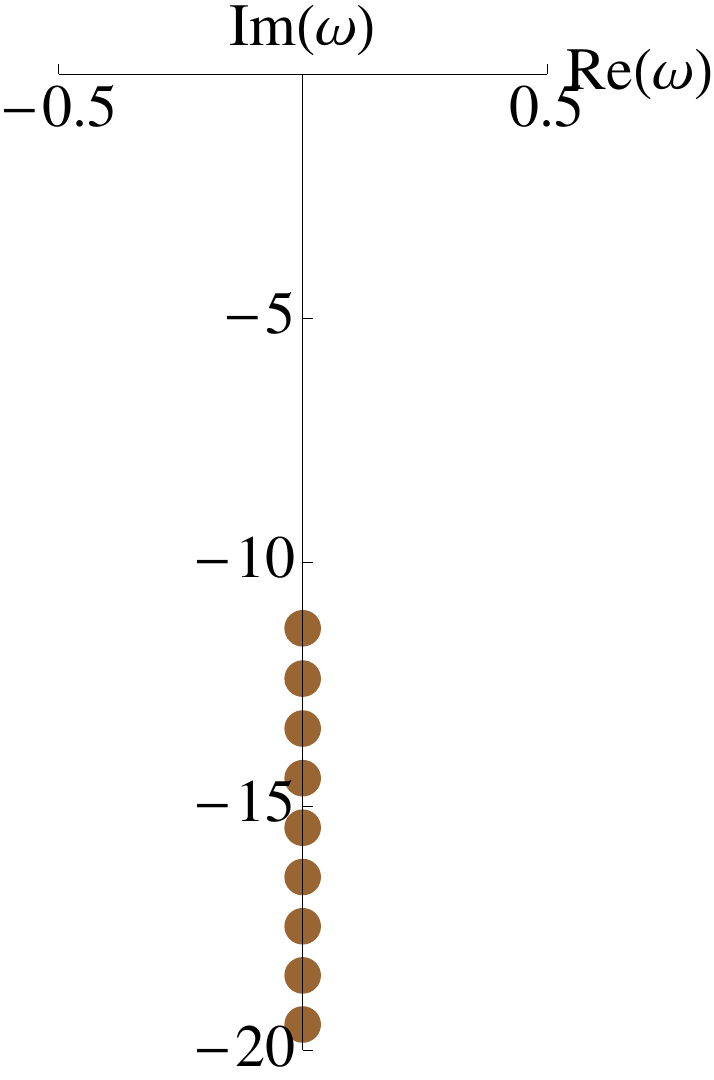}\hspace{0.8in}
\includegraphics[height = 1.8in]{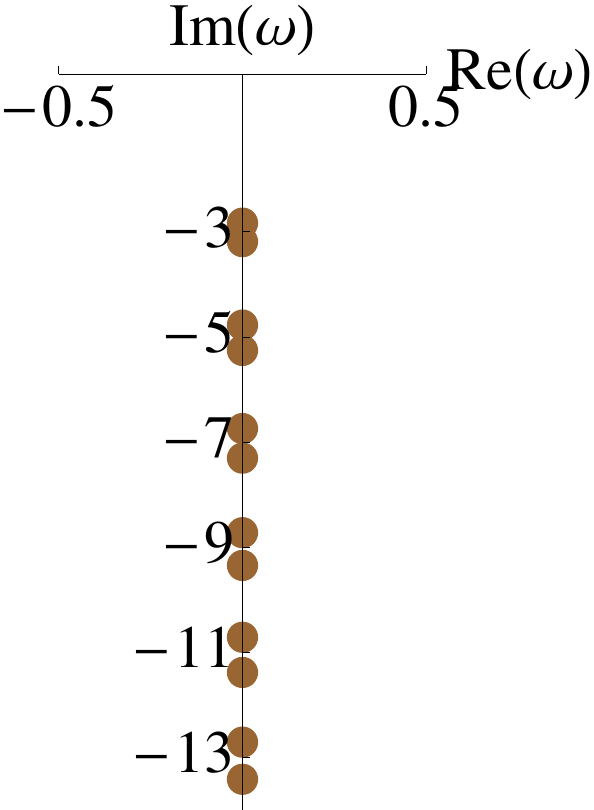}
\caption{\small{{\bf Left:} Location of quasinormal poles for $m_\varphi^2=7/4$ and $f_0=10$. {\bf Centre:} Quasinormal poles for 
$m_\varphi^2=475/4$ in the near AdS regime with $f_0=0.18$. Note that the lowest frequency is at $\omega \approx - i m_\varphi =-10.89i$. 
 {\bf Right:} Poles for $m_\varphi^2=475/4$ with $f_0=900$. The lowest mode is now at $\omega \simeq -3i$.  }}
\label{qnmmassive}
\end{figure}

 As the deformation $f_0$ is cranked up to arbitrarily large values, for any given scalar mass $m_\varphi$, we find that poles coalesce in pairs to yield singularities at
 \be
 \omega\,=\,-(2n+1)i\,,\qquad n=1,2,\ldots\,,  
 \ee
independent of $m_\varphi$.
This remarkable feature of the system is shown in figure \ref{qnmmassive}.
The reason for different operators exhibiting the same quasinormal frequencies can be traced back to the Schr\"odinger potential for the scalar wave equation \eqref{waveqn}. Taking the limit $z_0\to 0$ (equivalent to $f_0\to \infty$),
\be
V(z)\,\to\,\frac{4}{\sinh^2 z}\,-\,\frac{1}{\sinh^2 2 z}\,,
\ee
which is independent of $m_\varphi^2$. It is not {\em a priori} obvious that the $z_0 \to 0$ limit should commute with the $z\to 0$ limit required for computing boundary correlators and quasinormal frequencies. Our numerical computation of the quasinormal frequencies confirms that the large deformation limit leads to a universal result for the Green's functions of all scalar operators in this background. Note that our analytical results for the case $m_\varphi^2=0$ and numerical results for $m_\varphi^2<0$ are also consistent with this conclusion.

\section{Discussions: comparison with AdS black holes}
Various aspects of the retarded correlators we have found for  the deformed M2-brane CFT on dS$_3$ are strikingly similar to holographic thermal correlators obtained from AdS black hole geometries. The obvious similarities include the presence of quasinormal poles responsible for exponential decay of ($s$-wave) excitations as a function of de Sitter time. A somewhat more surprising feature is the quasinormal spectrum in the limit of infinite deformation parameter and the functional form of the retarded correlators in this limit, both of which  coincide with corresponding objects in the much simpler BTZ black hole (and higher dimensional topological AdS black holes with similar Penrose diagrams). 

These results are surprising since the dual theory is a relevant deformation of an intrinsically strongly coupled CFT, and there is no {\em a priori}  reason for the large deformation limit to be particularly simple.  In the event, the emergence of  equally spaced quasinormal frequencies (determined by odd integers) on the imaginary frequency axis, originating from the simple poles of the digamma function, must be seen to bear some relation to the shape of the spacelike singularity in the Penrose diagram in this limit. In particular, the Penrose diagrams (viewed side-on) in all these cases is a ``square'' and one is tempted to examine this resemblance in light of the simple geometric optics ideas of \cite{hoyos}. The suggestive link between the curvature singularity and quasinormal poles is underscored in our analysis of the massless bulk scalar ($\Delta=3$) where the appearance of one seems to be tied to the emergence of the other, even for arbitrarily small deformations. To add to this mix is the less obvious, empirical observation following from our numerical results for the $\Delta =5/2$ mode ($m_\varphi^2 = -5/4$) where the angle made by the line of quasinormal poles in the frequency plane appears to be related to the crunch coordinate $z_c$.

The observations above point towards a possible link between the crunching surface in the bulk geometry and singularities in frequency space correlators. Such a link was made precise for thermal correlators in the geodesic or WKB approximation \cite{shenkeretal, fl1, fl2, hoyos}, and the natural question is whether the same is possible for the crunching AdS models. Below, we point out the similarities and differences between the two situations (AdS black holes and crunching AdS duals), using the tractable AdS$_4$ model of this paper and the AdS$_5$-Schwarzschild  black hole as templates.

\paragraph{WKB and eikonal limits:}
Identification of interesting and subtle signatures of the AdS black hole singularity in frequency space correlators was possible chiefly due to  the existence of nontrivial spacelike geodesics (corresponding to purely imaginary frequencies in the boundary theory) that could probe the bulk singularity behind the horizon. Relevant aspects  of the analysis leading to this \cite{shenkeretal, fl1} are summarised in appendix \ref{appa}. Crucially, the Schr\"odinger potential (in tortoise coordinates) for a massive probe scalar in the black hole geometry splits in two pieces, each of which is sensitive to the singularity behind the horizon:
\be
V_{\rm BH}\,=\,q^2\,\left(1+r^2- \frac{\mu}{r^2}\right)\,+\, Q(r)\,,\qquad \mu \,=\,r_0(1+r_0^2)\,,
\ee
where $r_0$ is the Schwarzschild horizon radius and $q=\sqrt{d^2/4+m_\varphi^2}$ with $d=4$. It is easier to express $V_{\rm BH}$ in terms of the original Schwarzschild radial coordinate rather than the tortoise variable $z$. The first term in $V_{\rm BH}$ survives the geodesic limit $\omega, m_\varphi \to \infty$, whilst the second term $Q(r)$ is formally subleading in the WKB limit (since it does not scale with $m_\varphi$ or $\omega$). Note however that $Q(r)$ is dominant near the singularity at $r=0$ where $Q(r)\sim - 9\mu^2/4r^6 $. In terms of the tortoise coordinate, the near singularity behaviour of 
$V_{\rm BH}$ is
\be
\left. V_{\rm BH}\right|_{z\to z_c}\,=\, q^2\left(-\frac{1}{9\mu (z-z_c)^{2/3}}+\ldots\right)\,-\,\frac{1}{4 (z-z_c)^2}\,+\ldots
\ee
Here $z_c$ is the tortoise coordinate for the AdS$_5$ black hole crunch singularity:
\be
z_c\,=\,\frac{\pi}{2(r_0^2+r_1^2)}\left(i r_0\,+\,r_1\right)\,,\qquad
r_1=\sqrt{1+r_0^2}\,.
\ee
Importantly, the Schwarzschild potential in the formal WKB limit $q, \omega \to \infty$ (with $u=\omega/q$ fixed) continues to be divergent, presenting a repulsive potential for solutions with purely imaginary frequencies $(u^2 <0)$. Modes with large, negative  $u^2$ are thus presented with a WKB turning point near $z=z_c$. Therefore, whilst the leading singularity in the full potential $V_{\rm BH}$, scaling as $-\frac{1}{4}(z-z_c)^{-2}$, does not play any role in the WKB limit, the subleading singular terms provide a WKB or semiclassical turning point arbitrarily close to the singularity. It immediately follows that the WKB phase integral and therefore the WKB Green's function for such modes scales as 
\be
\tilde G_{\rm wkb}(u)\,\sim\,e^{-2 q\,|{\cal E}|\, z_c}\,,\qquad u\,=\,i {\cal E}\,,
\ee
and leads to an exponential decay of the frequency space Green's functions along the imaginary frequency axis \cite{fl1}. Note that $z_c$ is the complex Schwarzschild time taken by a null  geodesic to get from the AdS boundary to the singularity.

The Schwarzschild potential \eqref{waveqn} for the  crunching AdS geometry analysed in this paper differs from $V_{\rm BH}$ in at least one crucial aspect. Whilst it diverges at the crunch precisely as $-\tfrac{1}{4}(z-z_c)^{-2}$, the formally leading piece in the WKB limit, $V_{\rm wkb}\,=\,m_\varphi^2\, a^2(z)$, actually vanishes at the crunch. Therefore $z=z_c$ does not present a classical turning point to high (imaginary) frequency modes which, consequently, ``fall'' into the crunch. It  then becomes necessary to either  go beyond the leading order WKB approximation, or  to understand precisely what boundary conditions are natural for putative high frequency solutions at $z=z_c$ where $V(z)$ diverges.

A possible route\footnote{We thank Carlos Hoyos for drawing our attention to this possibility.}, distinct from WKB, is a straight high frequency or eikonal approximation which was employed in \cite{hoyos} to obtain an intriguing physical picture for the location of the high frequency quasinormal poles of AdS black holes. In particular, the eikonal approximation results in a ray optics description for high frequency, null waves which breaks down in the vicinity of the singularity at $z=z_c$. One may however, match the eikonal solutions to the actual solutions of the wave equation in the singular potential and consistently employ reflecting boundary conditions at the black hole singularity {\em and} the AdS boundary. The Penrose diagram thus acts as a reflecting cavity and the complex time delays suffered by the bouncing rays could be interpreted as light cone singularities in boundary correlators which lead to quasinormal poles satisfying (at high frequencies)
\be
\frac{{\rm Im}(\omega)}{{\rm Re (\omega)}}\,=\, \frac{ r_0}{r_1}\,=\,\frac{{\rm Im}(z_c)}{{\rm Re} (z_c)}\,.  
\ee
Whilst it is tempting to invoke a similar description to explain some of the properties of frequency space correlators we have encountered in this paper, a key conceptual difference between crunching AdS geometries and AdS black holes is the nature of the conformal boundary. While black holes have two conformal boundaries with time flowing in opposite directions, this is not the case for the crunching backgrounds. It is then unclear what meaning, if any, can be ascribed to geodesics bouncing off the singularity and reaching the boundary.

Nevertheless there are various concrete questions that  appear to be both puzzling and intriguing, and therefore worthy of future study. These include the nature of the singularities in the holographic correlator for  the massless bulk field (dual to a $\Delta=3$ operator) at arbitrarily small deformations $f_0 \ll 1$. We have found that the residue of the $n$-th quasinormal pole scales as $f_0^{2n}$ in the small deformation limit. This is also the limit in which the singularity just appears, and therefore the scaling of the residues with $f_0$ has a  potentially interesting physical origin. The behaviour of the phases of, or the angle made by the line of quasinormal poles for $m_\varphi^2 <0 $ and its relation to the location of the crunch singularity is perhaps the most intriguing of all questions, given what is already known for AdS black holes. In this paper we have steered clear of two important (and related) questions, both of which relate to the stability of the theory.  We have not computed correlators for operators with $1 \leq \Delta \leq 2$, and in particular for fluctuations of the scalar $\Phi$ which has a nontrivial profile and supports the solution. It would be interesting to look at the quasinormal poles for such operators and their motion in the complex frequency plane with increasing $f_0$, and uncover any direct signals of dynamical instabilities. Related to this is the fact that we have been able to dial the parameter $f_0$ from zero to infinity with impunity without inducing instabilities in ($s$-waves of) the scalar operators with $\Delta > 2$ since the poles reside in the lower half frequency plane. This brings us to a slightly puzzling feature not shared by relevant deformations yielding crunches in higher dimensions. In \cite{paper2}, where we analyze mass deformations of ${\cal N}=4$ SYM on $S^4$, dialling the deformation beyond a critical value forces the theory into a ``gapped'' Euclidean phase which results, upon analytic continuation, in a Lorentzian geometry without a crunch. This does not appear to be the case for the relevant deformation of the 3d theory we have analysed in this paper and it would be interesting to understand the origin of this difference.

 \acknowledgments 
We would like to thank Justin David, Tim Hollowood, Carlos Hoyos, and Soojong Rey for enjoyable discussions on various points related to this paper. SPK would like to thank Soojong Rey and The Center for Theoretical 
Physics, Seoul National University for hospitality and providing a stimulating environment while this work was being completed.
 This work was supported in part by the STFC grant awards ST/L000369/1 and ST/K5023761/1.

\appendix
\section{AdS-Schwarzschild revisited}
\label{appa}
It is useful to revisit the WKB analysis of holographic correlators in the Schwarzschild-AdS geometry which was originally performed in detail in \cite{fl1, fl2}.  The similarities and differences with the crunching-AdS models are both interesting and instructive.
The AdS$_5$-Schwarzschild geometry is given (in the global patch) by the metric:
\bea
&&ds^2\,=\,-f(r)dt^2\,+\,\frac{dr^2}{f(r)}\,+\,r^2 d\Omega_3^2\,,\\\nonumber
&& f(r)\,=\, 1\,+\, r^2\,-\, \frac{\mu}{r^2}\,,
\eea
where $\mu$ is the mass of the black hole. Following the useful notation introduced in \cite{fl1} 
\bea
\mu\,=\,r_0^2\, r_1^2\,,\qquad \qquad r_1^2\,=\, 1+r_0^2\,,
\eea
where $r_0$ is the radius of the black hole horizon.  In order to compute holographic correlators of some probe field $\varphi$ with mass $m_\varphi$ in the limit of high frequency and high mass (suppressing possible dependence on spatial/angular momenta), the problem is recast in Schr\"odinger form in the tortoise coordinate,
\be
z\,=\,\int_r^\infty\frac{dr}{f(r)}\,,
\ee
accompanied by a rescaling of the massive scalar field $\varphi$ and a separation of variables:
\be
\varphi\,=\,r^{-3/2}\,\sum_\ell\int d\omega\,e^{-i\omega t}\, Y_\ell(\Omega)\,\,\psi_{\omega,\ell}(z)\,.
\ee
The wave equation for $\varphi$ becomes a Schr\"odinger-like equation for the the radial mode $\psi_{\omega,\ell}(z)$. Suppressing the $\ell$-dependence for simplicity  and focussing attention on the $\ell=0$ mode, we have,
\be
-\psi_\omega''(z)\,+\, V_{\rm BH}(z)\,\psi_\omega(z)\,=\,\omega^2\,\psi_\omega(z)\,.
\ee
The potential term $V_{\rm BH}(z)$  is most easily understood in the original Schwarzschild radial variable,
\be
V_{\rm BH}(r)\,=\, f(r)\,\left[q^2-\frac{1}{4}\,+\,\frac{9\mu}{4r^4}\,-\,\frac{1}{4\,r^2}\right]\,.
\ee
where $q=\sqrt{\frac{d^2}{4}+m_\varphi^2}$ with $d=4$ for AdS$_5$.
The WKB limit is viewed as a high frequency and large mass limit. It is also possible to discuss a pure high frequency or eikonal limit, but this does not coincide with the geodesic approximation in general. The geodesic WKB limit is obtained by taking $\omega\,=\,q u$ with $q \gg 1$ whilst keeping the rescaled energy/frequency $u$ fixed.  In this limit the Schr\"odinger potential contains parametrically leading and subleading pieces:
\be
V_{\rm BH}\,=\,q^2\, f(r)\,+\, Q(r)\,.  
\ee
The solution to the leading order WKB problem and the resulting holographic Green's function is completely determined by the WKB phase integral 
\be
{\tilde G}_{\rm wkb}(u)\,\sim\,\exp{q Z(u)}\,,
\ee
where 
\bea
&&Z(u)\,=\,-2 \lim_{\Lambda\to\infty}\left({\cal I}_{\Lambda}(u)\,-\ln \Lambda\right)\,,\\\nonumber\\\nonumber
&&{\cal I}_\Lambda(u)\,=\,\int_{r^*(u)}^\Lambda dr\,\frac{\sqrt{f(r)-u^2}}{f(r)}
\eea
The lower limit $r^*(u)$ is the physical turning point for geodesics having real energies. For $u^2>0$, this physical turning point lies outside the horizon and approaches the AdS boundary as $u$ is increased. For complex $u$ this turning point must be analytically continued and its position carefully identified in order to distinguish it from other (complex) turning points of the WKB potential. Of particular interest is the case with $u^2 <0$ when the geodesics in question penetrate the horizon. 

Let us first express the WKB phase integral in a form which makes its evaluation more efficient. Moving to the new integration variables $x=r^2$, 
\be
{\cal I}_{\Lambda^2}(u)\,=\,\frac{1}{2}\int_{x_1(u)}^\Lambda dx\,\frac{\sqrt{(x-x_1)(x-x_2)}}{(x+r_1^2)(x-r_0^2)}\,,
\ee
where  $x_1 >0$ and $x_2<0$ for $u^2>0$:
\be
x_{1,2}\,=\,\frac{1}{2}\left[(u^2-1)\pm\sqrt{(u^2-1)^2 \,+ \,4\mu}\right]\,.
\ee
Although the integrals can be performed analytically, in order to develop a physical intuition it turns out to be very useful to consider the limit of large $u$, i.e. $u \gg 1$. In this limit, we have 
\be
x_1\approx\, u^2\gg 1\,,\qquad x_2\,\approx\,-\frac{\mu}{u^2}\,.
\ee
Therefore, in the limit of large positive $u^2$, the physical turning point approaches the boundary, while there also exists a complex turning point close to the singularity (approached from imaginary values of $r$), but whose contribution is subdominant, and the leading behaviour of the WKB phase integral is simply
\be
{\cal I}_{\Lambda}(u)\,\approx\,\int_{u^2}^{\Lambda^2} dx\,\frac{\sqrt{x-u^2}}{x^{3/2}}\,=\,\frac{1}{2}\log\left(\frac{\Lambda^2}{u^2}\right)\,.
\ee
This yields the expected high frequency behaviour ${\tilde G}_{\rm wkb}\sim \omega^{2\nu} $ expected from conformal invariance and dimensional grounds. However, the integral also contains interesting subleading corrections which can be seen either by direct analytical evaluation, or by moving to a different region in the complex $u$-plane where the subleading corrections also become easy to evaluate. 

For purely imaginary frequencies $u^2 <0$, the branch point $x_1$ is actually a complex turning point because $x_1<0$  which yields imaginary values of $r$. The physical turning point is inside the horizon and given by the branch point $x_2$. Setting $u\,=\,i  {\cal E}$ with ${\cal E} \in {\mathbb R}$, for large imaginary frequencies we obtain
\be
x_1\approx\, -{\cal E}^2\,,\qquad x_2\,\approx\,+\frac{\mu}{{\cal E}^2}\,.
\ee 
Now the WKB phase integral is 
\be
{\cal I}_{\Lambda}(u)\,=\,\frac{1}{2}\int_{x_2(u)}^{\Lambda^2} dx\,\frac{\sqrt{(x-x_1)(x-x_2)}}{(x+r_1^2)(x-r_0^2)}\,,
\ee
where $x_2 \ll r_1^2, r_0^2$ and the pole at the horizon falls on the contour of integration, and furthermore, unlike the situation with $u^2>0$, the integral does not simplify automatically even for ${\cal E}^2 \gg 1$.
\begin{figure}
\centering
\includegraphics[width= 2.8in]{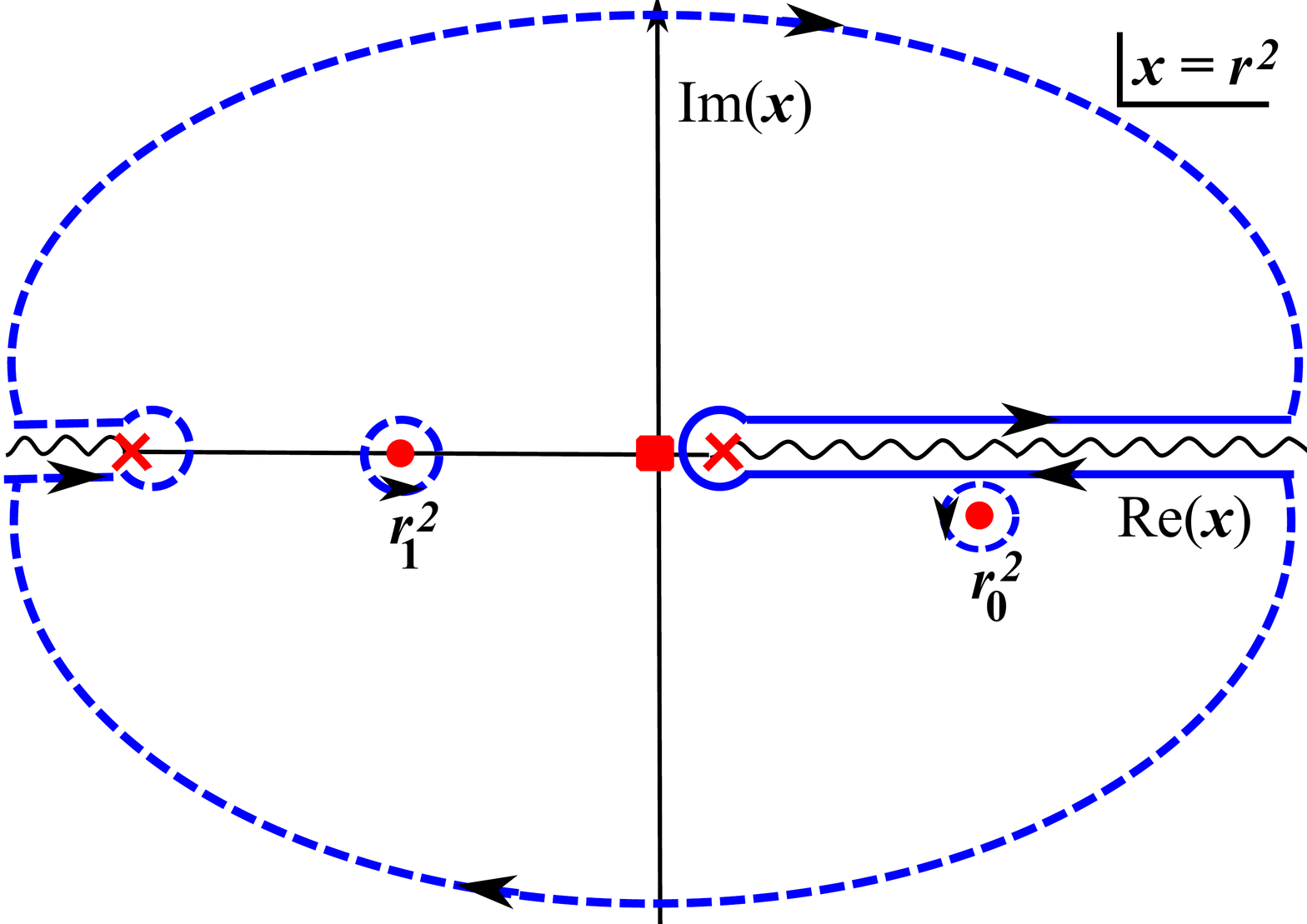}
\caption{\small{The WKB phase integral is performed along the branch }}
\end{figure}
However, given that the branch cuts of the integrand are square root branch cuts, the phase integral ${\cal I}_{\Lambda}$ can be expressed as one-half of the integral around the branch cut emanating from $x=x_2$, avoiding the pole at $x=r_0^2$ as shown in the figure. 
The integrand has a simple pole at $x=\infty$. As is usual the integration contour shown in the figure can be deformed smoothly so that wraps all the simple poles and the second branch cut emerging from $x=x_1$:
\be
{\cal I}_{\Lambda}\,=\,\frac{\pi}{2}\,{\cal E}\,\frac{ir_0 -r_1}{r_1^2+r_0^2}\,+\,
\frac{i\pi}{2}\,+\,\frac{1}{2}\ln\left(\frac{\Lambda^2}{{\cal E}^2}\right)\,.
\ee
The first term arise from the residues at the simple poles at the horizon and in the complex $r$-plane at $r^2 \,=\, -r_1^2$. There is an ambiguity in the sign of the real part of this term which should be fixed by a physical requirement. The second term is the contribution from the pole at infinity, while the all important last term which is required by conformal invariance, now arises from the complex branch point at $r^2 \approx - {\cal E}^2$. This explains the exponential falloff of frequency space correlators along the imaginary frequency axis observed by \cite{fl1}. It also explains the role played by the complex Schwarzschild time (or tortoise coordinate of the crunch) in the frequency dependence of the correlator.

\end{document}